%% file: main.tex
\documentclass[letterpaper,twocolumn,10pt]{article}
\usepackage{usenix-2020-09}
\usepackage{macro}

\usepackage{tikz}
\usepackage{amsmath}
\usepackage{tabularx}

\usepackage{filecontents}

\usepackage{authblk}
\makeatletter
\renewcommand\AB@affilsepx{, \protect\Affilfont}
\makeatother

\begin{document}

\date{}

\title{\bf \sysname{}: Self-scaling Stateful Actors For Serverless Real-time Data Processing}


\author[1]{\normalsize Le Xu\thanks{Contact author: Le Xu <le.xu@utexas.edu>}}
\author[1]{\normalsize Divyanshu Saxena}
\author[1]{\normalsize Neeraja J. Yadwadkar}
\author[1]{\normalsize Aditya Akella}
\author[2]{\normalsize Indranil Gupta}

\affil[1]{\normalsize University of Texas Austin}
\affil[2]{\normalsize University of Illinois at Urbana-Champaign}

\maketitle

\setlength{\belowdisplayskip}{2pt}

\begin{abstract}
We propose \sysname{}, a distributed stream processing service built atop virtual actors. \sysname{} achieves both a high level of resource efficiency and performance isolation driven by user intent (SLO). To improve resource efficiency, \sysname{} adopts a serverless architecture that enables time-sharing of compute resources among streaming operators, both within and across applications. Meanwhile, \sysname{} improves performance isolation by inheriting the property of function autoscaling from serverless architecture.  Specifically, \sysname{} proposes (i) dual-mode actor, an actor abstraction that dynamically provides orderliness guarantee for streaming operator during autoscaling and (ii) a data plane scheduling mechanism, along with its API, that allows scheduling and scaling at the message-level granularity. 
We show that through \sysname{}, an SLO-driven strategy could (i) improve performance isolation by increasing SLO satisfaction rate by 46\% while resources are shared between jobs, and (ii) achieve high resource efficiency by maintaining or even improving performance by using fewer resources through sharing at a fine granularity.  
 
\end{abstract}

\input{text/1-intro}

\input{text/2-motivation-v2}

\input{text/3-overview}

\input{text/4-actor}

\input{text/5-design}

\input{text/6-eval}
\input{text/7-related}

\input{text/8-conclusion}
\bibliographystyle{plain}
\bibliography{main}
\appendix
\input{text/9-appendix}

\end{document}

%% file: text/1-intro.tex
\section{Introduction}
\label{sec:intro}

Real-time data is known to bring unique provisioning challenges to dataflow engine design due to its unpredictable volume, velocity, and arrival patterns~\cite{birke2014big}. For today's distributed stream processing frameworks~\cite{toshniwal2014storm,sparkstreaming,flink_paper, heron,murray2013naiad,akidau2013millwheel,kreps2011kafka}, handling unpredictable real-time data while constantly satisfying various user-specified performance targets~\cite{stonebraker20058, floratou2017dhalion, henge}
requires elastic resource provisioning.

\begin{figure}
    \centering
    \includegraphics[width=.45\textwidth]{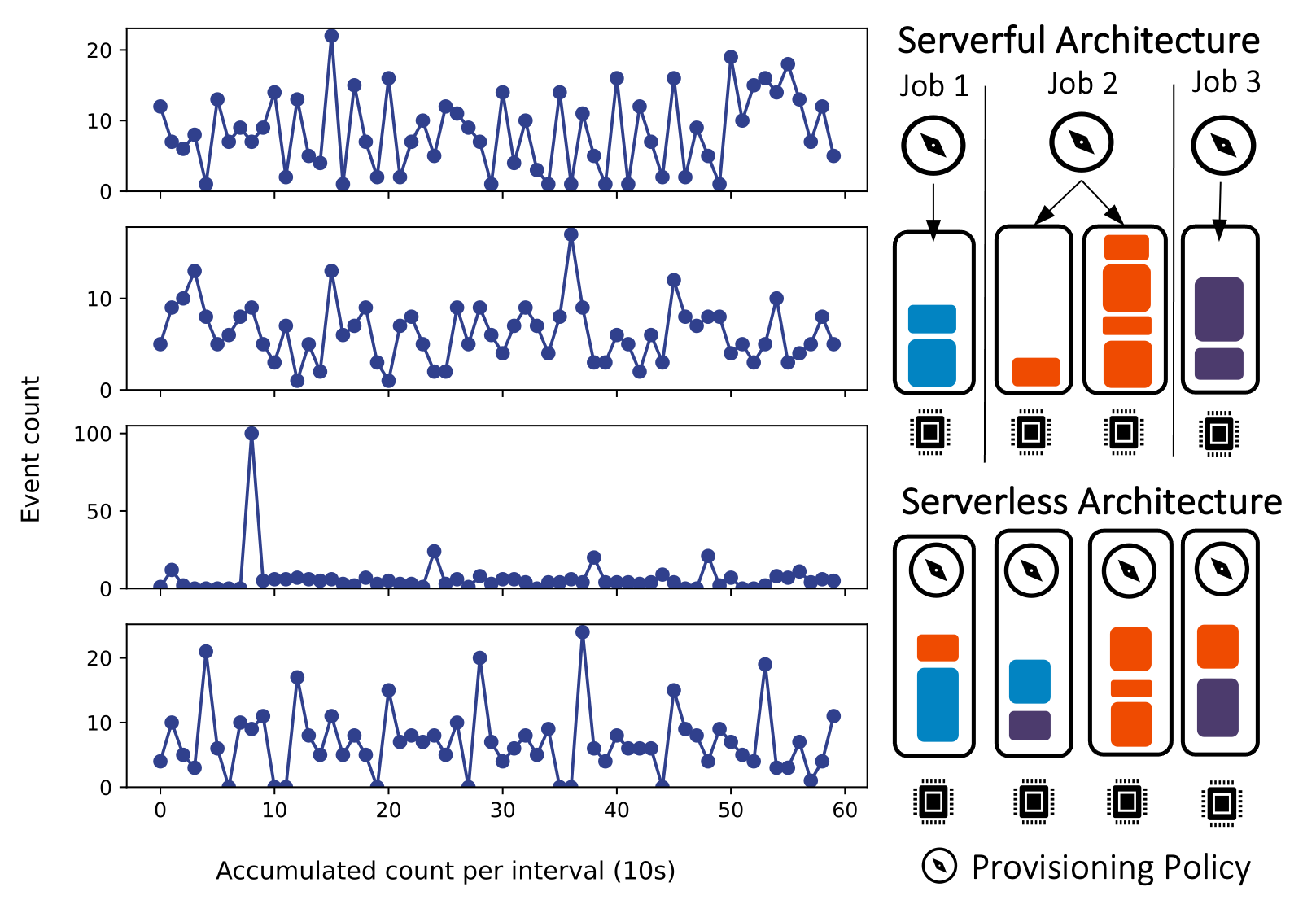}
    \caption{\textit{\small Left: Accumulated event volume (10s granularity) from four different Twitter stream~\cite{wang2014using}. Right: Operator scaling in serverful vs. serverless environment. 
    }}
    \vspace{-.2in}
    \label{fig:serverful-serverless-comparison}
\end{figure}

Existing state-of-the-art real-time dataflow engines use a fixed number of workers to run computational operators for different applications in an isolated fashion (called a serverful environment).
These state-of-the-art engines achieve elastic resource provisioning through reactive, job-level reconfiguration (Figure~\ref{fig:serverful-serverless-comparison}) --- processing pipelines need to be monitored at all times, and users perform diagnosis and generate new execution plans once performance violations or resource bottlenecks are detected\cite{floratou2017dhalion, kalavri2018three, del2020rhino}.
However, elastic resource provisioning remains a problem today, and users often have to pay up to 5$\times$ ``non-expert tax'' due to severe resource under-utilization even when providers support reconfiguration~\cite{wang2022non}. 

To demonstrate the challenge of elastic resource provisioning
, we show accumulated (10s) event volume from different Twitter traces in Figure~\ref{fig:serverful-serverless-comparison}.
The presence of short-term volume peaks and dips highlights an opportunity to use spare resources from streams experiencing a dip in input data to accelerate the execution of streams with bursts of input data. 
Today's systems are unable to tap into this opportunity as 
(i) they provision applications in a shared-nothing manner, preventing resources from being transferred between jobs efficiently, and
(ii) these systems use reconfiguration techniques to adapt to significant, long-term workload changes (e.g., diurnal workload changes) and perform reconfiguration actions at coarse granularity (even up to hours~\cite{wang2022non} for scale-in operations)

This paper explores a design of distributed stream processing services (DSPS) that enables resource sharing among hosted streaming applications. Our system, \sysname{}, adopts the design of serverless architectures~\cite{hendrickson2016serverless, bernstein2019serverless, kinesis} by modeling streaming operators as serverless functions that perform fine-grained time-sharing within the managed resources. 
While the serverless paradigm provides intrinsic benefits of high resource utilization through user-transparent function scaling, it also leads to the following challenges:

\noindent \textbf{Achieving fine-grained performance isolation:}
Conventional serverful streaming engines provision each application individually, and users continuously perform diagnoses to respond to the changing load and add/remove workers to the application.
However, adopting a serverless paradigm means systems do not reserve resources for individual jobs. Hence arbitrarily multiplexing requests targeting different applications could cause applications to fail to satisfy user intent (SLO). For instance, processing incoming requests in the FIFO manner could lead to latency-critical applications failing their target. Meanwhile, parallelizing many latency-insensitive requests is likely to cause performance degradation for other applications. 
Therefore, a DSPS needs to interpret performance targets and translate provisioning decisions by the scheduling strategies within a DSPS.
Meanwhile, provisioning decisions must be made frequently and carried out immediately to adapt to the input data's rate, shape, and distribution~\cite{fragkoulis2020survey}.

\noindent \textbf{Lack of support for auto-parallelizing streaming operators:}
While function autoscaling is natural to serverless architecture, it could lead to complications if we directly model a streaming operator as a serverless function: 
Firstly, many streaming operators are \textit{stateful} and frequently write to states – it is unclear how operator states should be accessed, maintained, and managed efficiently during the autoscaling process. 
Secondly, streaming operators need to meet various requirements for \textit{processing orders} of input events to produce correct and timely results, which could be broken easily by arbirtary parallel execution of incoming events. 
These challenges imply that a serverless DSPS should (i) natively manages in-memory state during the auto-scaling process, (ii) provide a way for the user to explicitly specifies ordering requirements and how states are partitioned and combined during the scaling process, and (iii) ensure correctness despite nondeterministic processing order introduced by parallelization. 
To address the first challenge, \sysname{} enables resource provisioning on the \textit{data plane} by invoking a scheduling policy that implements preset hooks on the execution path of each message. 
Data plane scheduling is critical for \sysname{} to minimize turnaround time for each scheduling decision so it can quickly respond to changing provisioning needs. 
\sysname{} provides a scheduling API for customizable scheduling policies to transfer SLO to real-time provisioning decisions (e.g., when and where to run each message).

To address the second challenge, 
\sysname{} provides a set of internal primitives used by scheduling policies to support automatic scale-out and scale-in of streaming operators. 
We introduce \textit{dual-mode actors} (\DMA{}) -- a virtual actor model~\cite{bernstein2014orleans} that switches between sequential and parallel execution modes to meet the ordered execution requirement of operators. 
Through \DMA{}, \sysname{}'s scheduling policy can (i) create a critical region where a streaming operator could execute messages in a single-threaded fashion and (ii) parallelize a streaming operator outside the critical region. \DMA{} enables \sysname{} to parallelize the streaming operator without violating the ordering requirement from the processing semantics. 

\sysname{} aims to provide the benefit of autoscaling and a ``hands-off'' experience to the end users of stateful, real-time stream processing applications promised by the serverless paradigm.
Essentially, \sysname{} provides opportunities for system designers to explore designs of custom scheduling strategies to provide fine-grained performance isolation in the resource-sharing environment. 
 \sysname{}'s contributions include 
(i) dual-mode actors and function partitioning API --- an internal mechanism provided by \sysname{} runtime that supports parallelization of the streaming operator while meeting their ordering requirements (Section~\ref{sec:dual-mode-actor}).
(ii) a data-plane message scheduler and scheduling API, designed for serverless DSPSes, that improves performance isolation through supporting customized, SLO-driven strategy (Section~\ref{sec:scheduling}).
(iii) a system prototype and demonstration of scheduling policies help DSPS to better satisfy user SLO despite workload unpredictability (Section~\ref{sec:evaluation}).
We show that \sysname{} serves up to 12\% more requests satisfying user-specified SLOs while using up to 30\% fewer resources compared to state-of-the-art approaches.

%% file: text/2-motivation-v2.tex
\section{Motivation}
\label{sec:motivation}



\subsection{The Need For Fine-grained Provisioning}

\begin{figure*}[!htp]
  \begin{tabular}[b]{cc}
      \subfloat[
        \textit{
        ~\label{subfig:social-network-pattern-hour}}
      ]{%
        \includegraphics[width=0.24\textwidth]{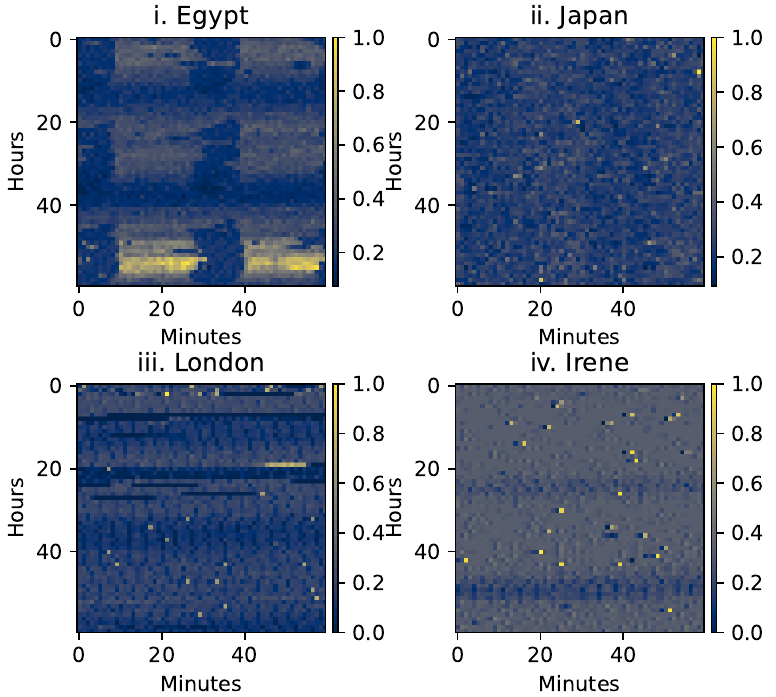}
      }
      \subfloat[
        \textit{
        ~\label{subfig:social-network-pattern-minute}}
        ]{%
      \includegraphics[width=0.24\textwidth]{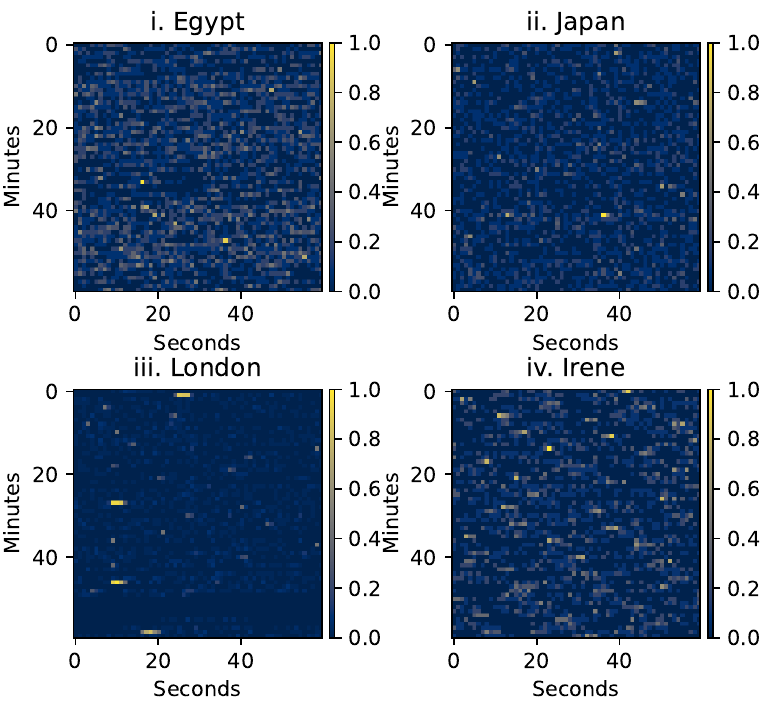}
      }
    
      &

      \begin{tabular}[b]{c}
        \subfloat[
          \vspace{-.1in}
          \textit{
          ~\label{subfig:log-interesting-event}. 
          }]{%
        \includegraphics[width=0.48\textwidth]{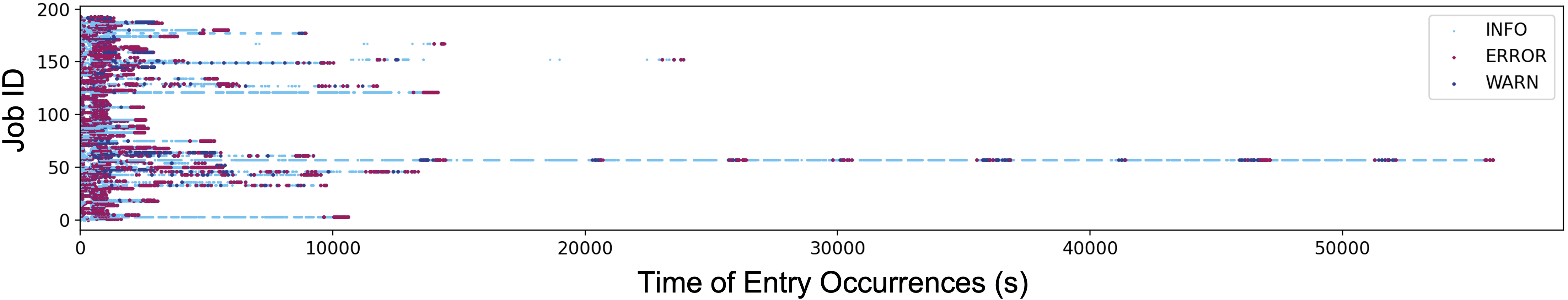}
      }
      \\
      \subfloat[
        \vspace{-.1in}
        \textit{
          ~\label{subfig:log-key-distribution}}
        ]{%
        \includegraphics[width=0.48\textwidth]{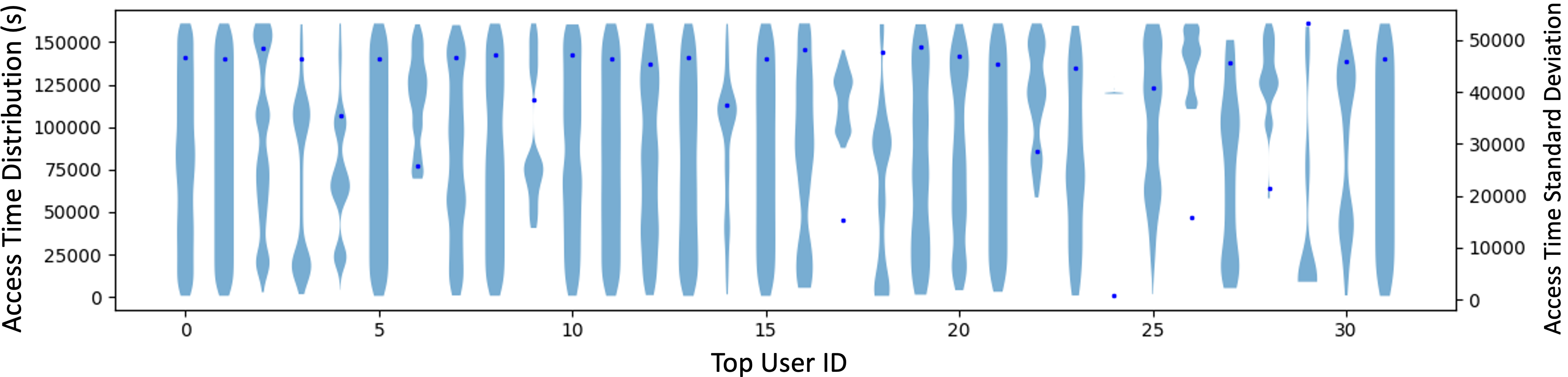}
      }
      \end{tabular}
  \end{tabular}
  \caption{\small \textit{
    \ref{subfig:social-network-pattern-hour} \ref{subfig:social-network-pattern-minute}: Normalized Twitter events volume during social events~\cite{wang2014using}: Eygyt Unrest, Japan Tsunami, London Riots, Hurricane. In \ref{subfig:social-network-pattern-hour}, each horizontal band shows the number of events at per minute granularity within an hour (normalized in [0,1]). \ref{subfig:social-network-pattern-minute} shows the same metric at the granularity per second. \ref{subfig:log-interesting-event}: Each band represents a job, and the times of events with different levels of criticality occur~\cite{he2020loghub}. \ref{subfig:log-key-distribution}: Each vertical band shows the number of accesses of 32 top users that access the cluster during a 44-hour period ~\cite{reiss2011google}. The blue dot shows the standard deviation of these accesses. 
  }}

  \label{fig:log-event-pattern}
\end{figure*}
    
\begin{figure}[htp]
  \centering
  \includegraphics[width=.5\textwidth]{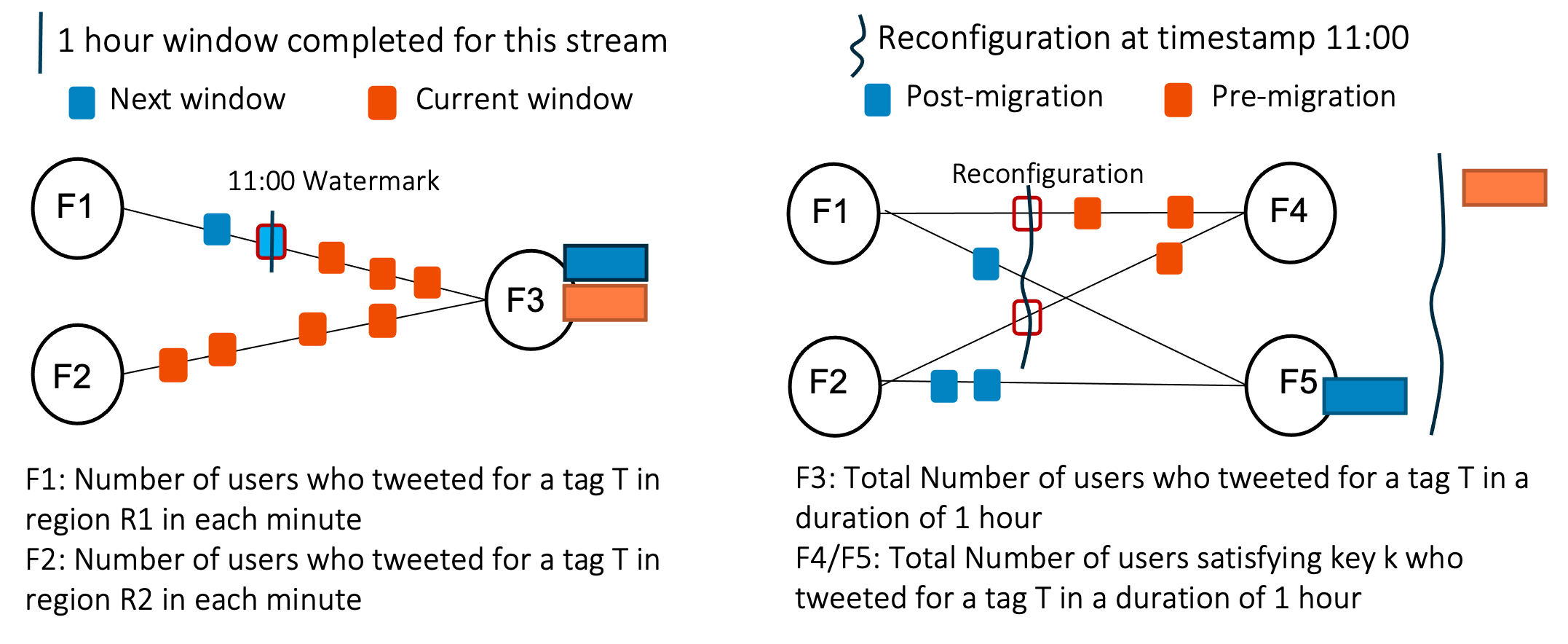}
  \caption{\small \textit{Sensitivity of real-time dataflow to event processing order. Left: Function $F3$ performs window aggregation and receives a watermark\cite{akidau2015dataflow} event signaling the completion of window. Right: $F1$ and $F2$ initiate a barrier synchronization on $F4$ and migrate its processing states to $F5$.}}
  \label{fig:ordering-scenario}
\end{figure}

\noindent\textbf{Inherent workload variability: } 
Figures~\ref{subfig:social-network-pattern-hour} and~\ref{subfig:social-network-pattern-minute} show the distribution of tweet post events during trending social events.
Figures~\ref{subfig:log-interesting-event} and~\ref{subfig:log-key-distribution} show two sets of cluster traces used for anomaly detection or reporting critical cluster events. 
 
We note the following traits of real-time data streams, common in various usage scenarios: 
\newline i).~\textit{Short-lived workload changes exist:} Figures~\ref{subfig:social-network-pattern-hour} and~\ref{subfig:social-network-pattern-minute} show input spikes (e.g., Figure~\ref{subfig:social-network-pattern-hour} (iii)) and dips (e.g., Figure~\ref{subfig:social-network-pattern-minute} (i)) with more than 2$\times$ load short-lived (lasting for seconds to minutes) changes. 
In some scenarios (e.g., Figure~\ref{subfig:social-network-pattern-minute} (iii)), we observe that the input stream receives large amounts of data during a short period before remaining idle for the rest of the time. The same traits could also be observed in Figure~\ref{subfig:log-key-distribution}, where the popularity of a particular data attribute (e.g., user) in some streams remains constant throughout (user 0). In contrast, some user (user 24) becomes extremely popular for a short period but remains inactive for the rest of the time. 
\newline ii). \textit{Not all changes are predictable:} We observe some predictable workload changes in Figures~\ref{subfig:social-network-pattern-hour} and~\ref{subfig:social-network-pattern-minute}, implying an opportunity for providers to reserve or de-allocate resources with high accuracy. 
However, we also observe less predictable load changes, with load peaks and dips occurring without clear periodic patterns. Similar patterns could be observed in from Figure~\ref{subfig:log-interesting-event}: for users who monitor errors in their jobs, the log entry containing ``ERROR'' could occur at any time. Such events need to be processed immediately to avoid delays in failure handling.


\noindent\textbf{Differences in application needs: } Streaming applications may have varying performance, and in turn resource, requirements.
Applications typically specify these requirements as  latency~\cite{kalyvianaki2012overload,zhang2018awstream,lohrmann2015elastic, heinze2014latency, henge}, throughput~\cite{henge, stream_ingestion_synapse}, and fairness~\cite{kalyvianaki2016themis} goals.
The ability to intelligently translate performance needs into provisioning decisions is critical.
\noindent\textbf{Takeaways:}  
Our observations introduce design challenges from the following aspects:
\colorcircled{\scriptsize R1} \textit{resource sharing:} To achieve resource efficiency, a DSPS needs to share resources across jobs while meeting their performance reuqirements. To do so, a DSPS needs to accommodate load bursts by swiftly sharing resources from other jobs. 
\colorcircled{\scriptsize R2} \textit{provisioning granularity:}  To adapt to frequently changing provisioning needs within and across applications, a DSPS has to opt for \textit{lightweight} \textit{fine-grained} resource provisioning. 
\colorcircled{\scriptsize R3} 
\textit{translating application needs: }
To make provisioning decisions that multiplex resources among applications, a DSPS needs to understand both the \textit{service level objectives} (SLOs) and the \textit{restrictions} imposed by processing semantics of an application. 

Processing semantics of incoming requests drive the decisions about \textit{where and how to perform scaling}. Event ordering is critical to stream processing applications: Figure~\ref{fig:ordering-scenario} shows two examples where stream processing applications execute event \textit{barriers} before all pending events and after all its past event dependencies. 
A DSPS that performs autoscaling must respect the orderliness of events when deciding {\em what to process next} and \textit{whether to parallelize} consecutive events that target a specific operator.

\vspace{-1em}

\subsection{Existing Approaches}
\label{subsec:existing-approaches}

\subsubsection{Cloud-based DSPSes}
\noindent\textbf{Dynamic resource provisioning for a DSPS:}
Today's DSPSes~\cite{aws_kinesis, kinesis, google_dataflow} and distributed streaming engines that supports multi-tenancy~\cite{toshniwal2014storm, sparkstreaming, murray2013naiad, flink_paper, heron} are primarily designed for a serverful cloud
and therefore do not natively support \colorcircled{\scriptsize R1}. Resources are dynamically acquired and released through \textit{reconfiguration} that could re-partition operators, change processing logic, and switch dataflow plan~\cite{fragkoulis2020survey, roger2019comprehensive} in an automatic~\cite{gedik2014elastic, li2015supporting, floratou2017dhalion, kalavri2018three} or on-demand~\cite{castro2013integrating,xu2016stela} manner. 
Reconfiguration is the state-of-the-art method to mitigate long-term workload changes (e.g., diurnal or bi-modal pattern). 
While effective, it is a costly~\cite{mai2018chi, hoffmann2019megaphone, gu2022meces} way of resource planning triggered by the \textit{dataflow controller}~\cite{floratou2017dhalion, mai2018chi, hoffmann2019megaphone}, and it typically takes several seconds to an hour~\cite{wang2022non}. 
Dynamic provisioning that relies only on reconfigurations prevents a DSPS from fully satisfying \colorcircled{\scriptsize R2}. 
While reconfiguration is the state-of-the-art method to mitigate long-term workload changes, fully exploiting workload and resource availability requires us to explore an alternative architecture beyond a serverful design. 

\noindent\textbf{Support for multi-tenant performance isolation:} Most existing SPEs perform resource scaling on a \textit{single} application. These solutions are feasible under the serverful environment, as concurrently running applications get isolated resources and do not interfere with each other. 
Henge~\cite{henge} targets \colorcircled{\scriptsize R3} through resource reconfiguration/reduction/reversion at operator granularity with a feedback loop based on cluster performance metrics derived from SLOs. However, its periodic feedback loop (10s interval) can only adapt to long-term workload changes and fails to satisfy \colorcircled{\scriptsize R2}. 

\vspace{-1em}
\subsubsection{Adopting Serverless Design}
The serverless paradigm, by default, satisfies~\colorcircled{\scriptsize R1} through fine-grained provisioning and allows sharing of resources among applications. However, today's serverless systems are insufficient from the following perspectives. 

\noindent \textbf{SLO-driven fine-grained provisioning:}
Performance-driven resource provisioning for the serverless paradigm has been a focus of recent works focusing on provisioning strategies and mechanisms meeting various performance targets (e.g., latency~\cite{singhvi2021atoll, jia2021nightcore}, throughput~\cite{schuler2021ai}, cost~\cite{gupta2020utility} or customized targets~\cite{tariq2020sequoia}) of serverless applications. However, most of these rely on a (semi-) control plane scheduling for provisioning decisions through feedback loops. The communication between the data and control plane prevents systems from making provisioning decisions on every function invocation (failing \colorcircled{\scriptsize R2}). Wukong~\cite{carver2020wukong} discusses a decentralized scheduler design that performs independent autoscaling on each function within a serverless DAG. However, it targets IO minimization rather than satisfying SLO.


\noindent \textbf{Semantic awareness:} 
Streaming applications cannot fully enjoy the benefit provided by serverless architecture today due to the following properties:

  \noindent\textit{(i) Write-heavy state accesses:} Many streaming applications are \textit{stateful} and write to states upon receiving every input. They are intolerable to the cost of remote state access and require in-memory managed states. 
  Deploying stream processing pipelines today requires a user to divide processing pipelines into stateful and stateless stages. Only stateless stages are deployed through serverless functions~\cite{streaming-serverless-app, serverless_event_processing_confluent} and can be auto-scaled. This is because serverless offerings, such as~\cite{awslambda}, require functions to manage state externally in the cloud storage, which makes them unsuitable for streaming applications. 
  
  \noindent\textit{(ii) Sensitivity to processing order: } Many streaming operators rely on the notion of ``timestamp''~\footnote{Timestamps could be physical wall clock time, or logical time that signify the progress of a data stream.} to trigger their actions. 
  These streaming operators need to process \textit{critical events} as barriers (e.g., watermark~\cite{akidau2015dataflow}, punctuation~\cite{chandramouli2014trill}, etc.) between their \textit{causally dependent} past events (bearing timestamps that are earlier to the critical events) and \textit{causally pending} future events.
  Ensuring these orders is important to (i) produce the correct result (Figure~\ref{fig:ordering-scenario} left) and (ii) carry out critical functionalities, including checkpointing~\cite{chandy1985distributed,carbone2017state} and reconfigurations~\cite{mai2018chi, mao2021trisk}(Figure~\ref{fig:ordering-scenario} Right). 
  Modeling a streaming operator as a serverless function poses a significant restriction on autoscaling to existing serverless frameworks, making them unable to satisfy \colorcircled{\scriptsize R3}. 
  Events can be executed in a nondeterministic order in parallel during autoscaling. Thus blindly parallelizing events that target an operator could lead to causally inconsistent results (e.g., loss of updates, replaying previously executed input during recovery, etc.) 
  To mitigate this challenge, a DSPS needs to create distributed barriers dynamically while a function is being parallelized.
  Netherite\cite{burckhardt2022netherite} adopts a two-phase commit protocol to generate a critical region for incoming messages. However, this is insufficient to support barriers as it cannot reason about ordering between events.
  HydroCache~\cite{wu2020transactional} and Boki~\cite{jia2021boki} propose a mechanism to ensure causal consistency for transactions for replicated function states through a storage layer (via logs/KVstore).
  However, a critical event for \sysname{} could be targeting a stateless function, making it necessary to explore a \textit{storage-free} coordination mechanism (we discuss this further below).  
\subsection{The Case for Virtual Actors}


\sysname{} adopts \textit{the virtual actor model}~\cite{bernstein2014orleans, dotnet_orleans} and maps each DAG operator (function)~\footnote{A streaming operator maps to a function and we use these terms interchangeably.} to a virtual actor.
This virtual actor model provides the following benefits: 
i). \textit{addressable functions with exclusive states:}
Addressable functions allow \sysname{} to deliver an invocation directly from one function to another without needing an external service such as a function coordinator. Therefore, \sysname{} can keep function states in memory without constantly tracking and transferring function states. 
ii). \textit{runtime-managed actor instances:} Unlike the original actor model, a virtual actor model manages actor instances through \textit{automatic instantiation} and \textit{location transparency}~\cite{bernstein2014orleans}. An actor is only \textit{instantiated} (assigning to the executor thread) when it receives a message to execute. 
An idle actor is removed from the executor (satisfying~\colorcircled{\scriptsize R1}) and the runtime stores its states until the actor is instantiated again (possibly at a different location in the cluster)\footnote{While this model allows states to be partitioned and (partially) offloaded to remote storage under memory constraint or through checkpointing, in this work we discuss in-memory function states. }.
The runtime has complete control over \textit{when} and \textit{where} to instantiate an actor, allowing \sysname{} to perform message-level resource provisioning (\colorcircled{\scriptsize R2}).


\begin{figure}[htp]
  \centering
  \includegraphics[width=.45\textwidth]{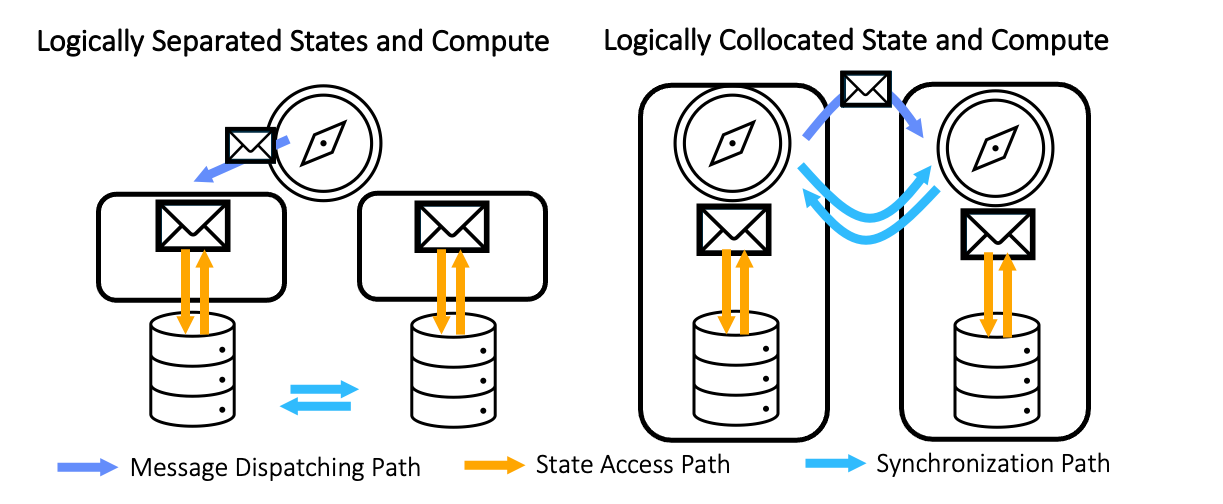}
  \caption{\small \textit{Logical diagrams of two designs: logically separated function compute and states and logically collocated function compute and states. }}
  \label{fig:arch-comp-scenario}
\end{figure}   
Cloud services adopt one of the two designs depicted in Figure~\ref{fig:arch-comp-scenario}. 
Most of today's serverless frameworks adopt the \textit{data-shipping} paradigm (Figure~\ref{fig:arch-comp-scenario} left):
It decouples function execution from state management. The function states are synchronized, persisted, and updated passively through external storage components (cache/log/KV store). The scheduler makes reactive decisions on where to execute a function (e.g., maximize state locality~\cite{sreekanti2020cloudburst}).
This `logical disaggregation with physical colocation' design is adopted by many recent proposals on supporting stateful functions~\cite{sreekanti2020cloudburst, burckhardt2022netherite, zhang2020fault, jia2021boki}. 

Through virtual actors, \sysname{} adopts a \textit{compute-shipping} paradigm that manages the functions \textit{natively along with} their states. This design greatly simplifies \sysname{}'s architecture: 
i). Even stateless streaming operators require ordering guarantees which would require the data-shipping paradigm to explicitly enable coordination through function states. This is unnecessary and \sysname{} can avoid them by a general mechanism.
ii). The write-heavy nature of stream processing applications and the fine-grained provisioning scheme by \sysname{} relies heavily on frequent updates to the function's computational states and scheduling states (e.g., routing information~\cite{castro2013integrating}).
Avoiding unnecessary overhead of state synchronizations requires complex coordination mechanisms between the function scheduler and storage layer in the data-shipping paradigm. Using a function-shipping paradigm helps \sysname{} combine all coordination decisions into a single entity.

%% file: text/3-overview.tex
\section{Overview}
\label{sec:overview}



\sysname{} models a data stream application as a DAG of user-implemented event-driven functions. Each function internally maps to a virtual actor.  
Once a dataflow job is submitted, \sysname{} registers these functions on a selection of \sysname{} workers.
Each function of a DAG in \sysname{} is associated with a unique \textit{function address}. A function invokes another function by sending \textit{messages} to the target function's \textit{mailbox}.
Each function address, by default, maps to a \textit{\sysname{} worker} in the cluster, who manages the function's mailbox. 
Each \sysname{} worker could contain mailboxes of different functions. 

\begin{figure}[htp]
  \centering
  \includegraphics[width=.5\textwidth]{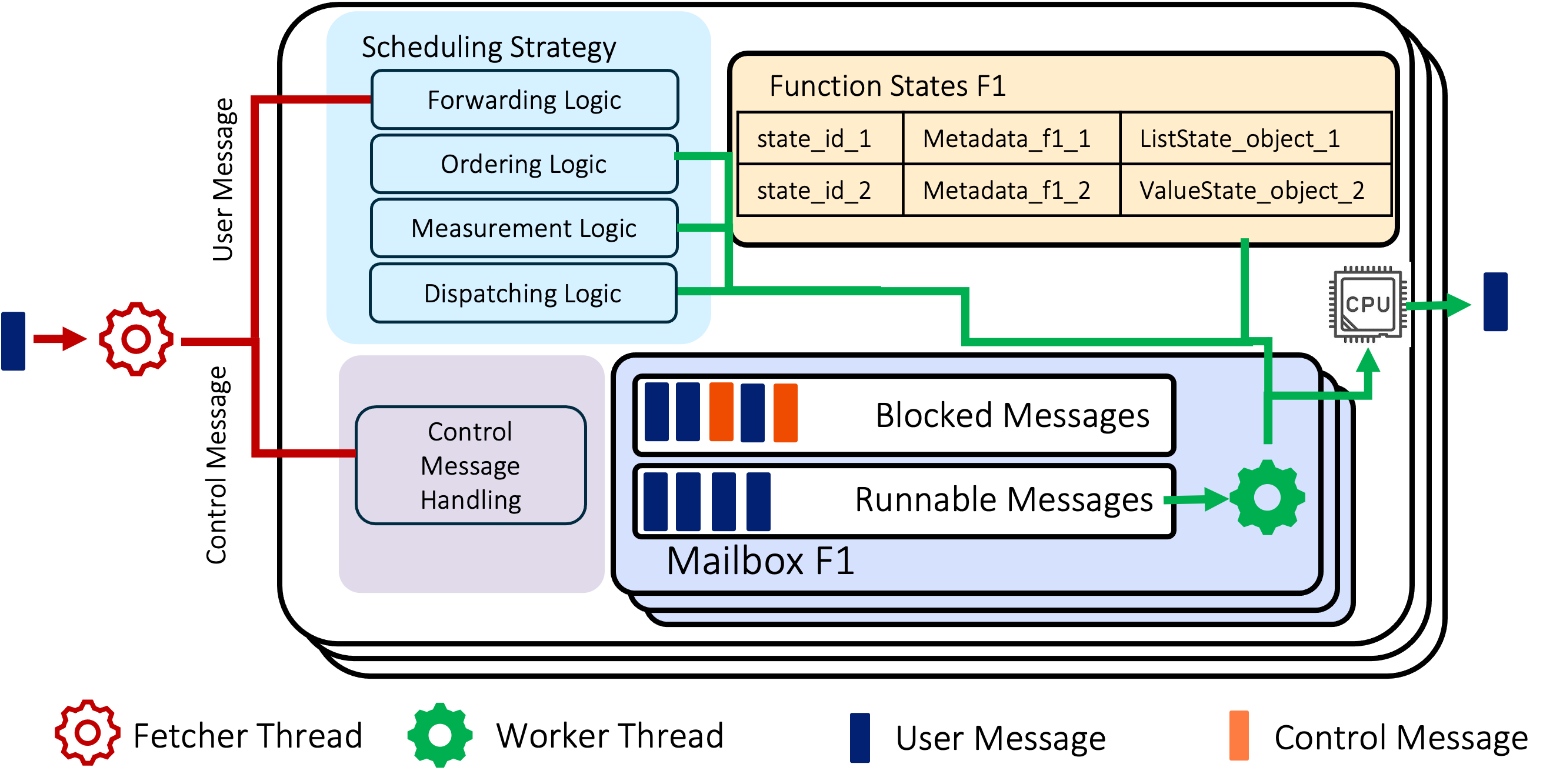}
  \caption{\textit{\small \sysname{} Runtime.}}
  \vspace{-.1in}
  \label{fig:architecture}
\end{figure}

Figure~\ref{fig:architecture} shows the logical diagram of \sysname{} components. \sysname{} runs multiple workers simultaneously, each with a Flink~\cite{apache_flink} instance running as an underlying message processor. Each worker consists of a fetcher thread and a worker thread. The fetcher thread receives all incoming messages and then enqueues them to the function mailbox. A worker thread runs a processing loop to execute messages from mailboxes that contain messages. 
\sysname{} messages can be either \textit{user messages} generated by the user functions or \textit{control messages}, special messages introduced by \sysname{} required that support dynamic function scaling (Section~\ref{sec:dual-mode-actor}).


\noindent\textbf{Fine-grained provisioning in \sysname{}:} 
\sysname{} enables provisioning at a per-message granularity by lowering scheduling decisions from the control plane to the data plane.
\sysname{} pre-sets a set of hooks along the execution path of each message. It exposes these hooks as a scheduling API that can be used to implement customized scheduling policies with a variety of scheduling behaviors (Figure~\ref{fig:architecture}) (e.g., message forwarding and ordering, profiling runtime statistics, selecting target workers, etc.). 
The scheduling policy has a complete view over all ready-to-run messages (possibly targeting functions that belong to different applications) within the worker. 
It also has complete control of selecting the next message to execute and deciding whether a message should be re-assign (i.e., \textit{forward}) to a different worker under load spikes.
Through this data-plane scheduling mechanism, \sysname{} could respond to load changes quickly without the overhead of communicating with an external scheduler. 
This message-based forwarding also allows \sysname{} to support implicit scale-out and scale-in by simply scheduling messages on different workers.

\begin{figure}[!ht]
  \centering
  \includegraphics[width=.45\textwidth]{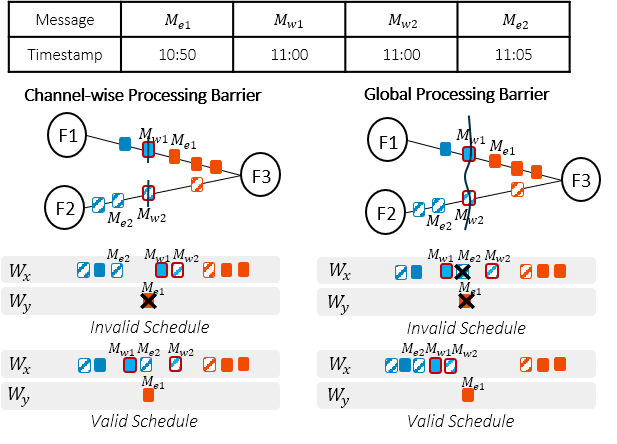}
  \caption{\textit{\small Autoscaling leads to arbitrary execution order: we show valid and invalid schedules as a timeline of messages for both scenarios.} 
  }
  \vspace{-.1in}
  \label{fig:event-ordering-example}
\end{figure}

\noindent\textbf{Parallelizing streaming operators: }
As described previously, scaling an operator dynamically by dispatching messages to parallel workers lead to nondeterministic processing order. 
Streaming applications rely on critical events to trigger execution (e.g., updating stream progress) and snapshotting (e.g., checkpointing and reconfigurations)~\footnote{In \sysname{} these events can be inserted by user and by scheduling policies.}. 
For streaming applications, preserving the processing ordering of these critical events is crucial to ensure the correctness of the result. 
Given three functions running on two workers ($W_{x}$ and $W_{y}$) and receiving incoming messages timed between 10:50 to 11:10, Figure~\ref{fig:event-ordering-example} shows two example scenarios:


\noindent\textbf{(i) Channel-wise Processing Barrier:} Figure~\ref{fig:event-ordering-example} (left) shows hourly window aggregation operators with two upstream $F1$, $F2$. At 11:00, the window closes and an output is generated based on all input from 10:00 to 11:00 (orange events).
At time 11:00, both upstream operators send watermark (or punctuation) events $M_{w1}$, $M_{w2}$, which signal that no items before 11:00 will arrive from the sender operator after the watermark message is seen (a typical approach adopted by existing streaming engines~\cite{akidau2015dataflow}), and the downstream operator $F3$ can generate an aggregated result based on all messages between 10:00 to 11:00.
In this scenario, scheduling message $M_{e1}$ from worker $W_{x}$ to $W_{y}$ could lead to a lost update if watermark $M_{w1}$ is executed on $W_{x}$ before executing $M_{e1}$ was executed on $W_{y}$.
A correct runtime here should ensure that the watermark message should be processed as a \textit{channel-wise processing barrier} by pausing the watermark message $M_{w1}$ (and all incoming messages from $F1$). All earlier messages from $F1$ should complete before the barrier (whether on $W_{x}$ or $W_{y}$), and no messages later than the barrier ($M_{w1}$) from $F1$ should be executed before $M_{w1}$.

\noindent \textbf{(ii) Global processing barrier:} Another typical streaming operation is that of a \textit{distributed global snapshot}, where all upstream send critical events to a downstream function. The required execution order imposes the constraint that messages must be paused from all upstreams.
This type of barrier is typically adopted by streaming engines to perform functionalities that require global synchronization, such as checkpoint and reconfiguration, etc.~\cite{carbone2015apache, mai2018chi}: In Figure~\ref{fig:event-ordering-example} (right), a valid schedule in the left figure would cause message $M_{e1}$ to be excluded from the snapshot and $M_{e2}$ to be incorrectly included in the snapshot.
In this scenario, the runtime should ensure that the control message should be processed as a \textit{global processing barrier}: All messages (orange) that arrived before 11:00 should be processed by parallelized $F3$ before any of the barrier messages, and all later messages from any upstreams should be processed after the barrier.



Beyond the complexity of execution order, scaling a stateful function to multiple workers will result in the partial function state being updated simultaneously over these workers. These partial function states must be combined dynamically when the critical events are present (e.g., when a watermark event triggers the streaming operator to produce output).
It is runtime's responsibility to exploit opportunities of parallelizing actors by scheduling messages to different workers while also providing the illusion of preserving single-threaded processing semantics when critical events are processed.

\sysname{} achieves this by proposing the abstraction of dual-mode actor (\DMA{}), wherein an actor is a logical single-threaded instance but physically disaggregated at multiple workers.
\DMA{} supports the use of \textit{shared leases}, where the \textit{lessor instance} of an actor (function) shares its lease with \textit{lessee instances} of the same actor on other workers. 
By default, a function runs on a single actor instance (called the \textit{lessor instance}), and the scheduling strategy can perform autoscaling on a function by assigning message(s) to another worker. This creates a shadow instance of the actor with shared lease (which we call a \textit{lessee instance}). 
Instances can be added or removed from these leases by using specific protocols - thus, allowing seamless switches between sequential and parallel modes of operation. Section~\ref{sec:dual-mode-actor} describes the design of \DMA{} and the associated primitives needed to achieve the same.

%% file: text/4-actor.tex
\section{\sysname{} Dual Mode Actors}
\label{sec:dual-mode-actor}

Here,
we specify the design of the \dma{} actor and describe the protocol it uses to seamlessly switch between the two modes of execution in Section~\ref{subsec:dma}. 
\sysname{} proposes a novel \DMA{} protocol to switch between two execution modes transparently: \textit{\PARALLEL{}}, where messages directed to a function can be executed in parallel and \textit{\SEQ{}}, where incoming messages are executed in a single-threaded fashion.
Next, we describe the trigger for the protocol (which we call \texttt{SYNC} program) in Section~\ref{subsec:sync-program}. Note that we relegate description of the API for the users to specify the processing and state access semantics in Section~\ref{subsec:user-api}.



\subsection{Dual Mode Actor Protocol}
\label{subsec:dma}



We denote messages that include critical events as \textit{critical messages}, and they act as \textit{barriers} that the runtime must respect.
We formalize the definition of critical messages and barriers as follows:
\newline\noindent \textbf{\textit{Critical Messages:}}~\textit{Critical messages are special messages that require a \SEQ{} mode of execution. Critical messages have other messages as dependencies and become dependencies for subsequent messages. Formally, a critical message $CM$ has an associated dependency set $\mathcal{D}_{CM}$ and an associated pending set $\mathcal{P}_{CM}$. All messages $m_{d} \in \mathcal{D}_{CM}$ must be executed before $CM$ is executed and $CM$ needs to be executed before all messages $m_{p} \in \mathcal{P}_{CM}$.}
\newline\noindent \textbf{\textit{Barriers:}}~\textit{A barrier $B$, is a set of critical messages that must be processed together. Given the set of critical messages $B = \{CM_{i}\}$, $B$ has an associated dependency set $\mathcal{D}_{B}$ (pending set $\mathcal{P}_{B}$), which is the union of dependency sets (pending sets) of all its constituent critical messages. All messages $m_{d} \in \mathcal{D}_{B}$ must be executed before any of the critical messages in $B$ can be processed. Further, message $m_{p} \in \mathcal{P}_{B}$ should be processed after all critical messages in $B$ have been processed.}

\begin{figure*}[!htp]
  \subfloat[\textit{ \DMA{} in action: Sending a critical message from upstream lessor instance $U_{L}$ to parallel downstream actor $D$.~\label{fig:barrier-protocol}}]{%
    \includegraphics[width=0.59\textwidth]{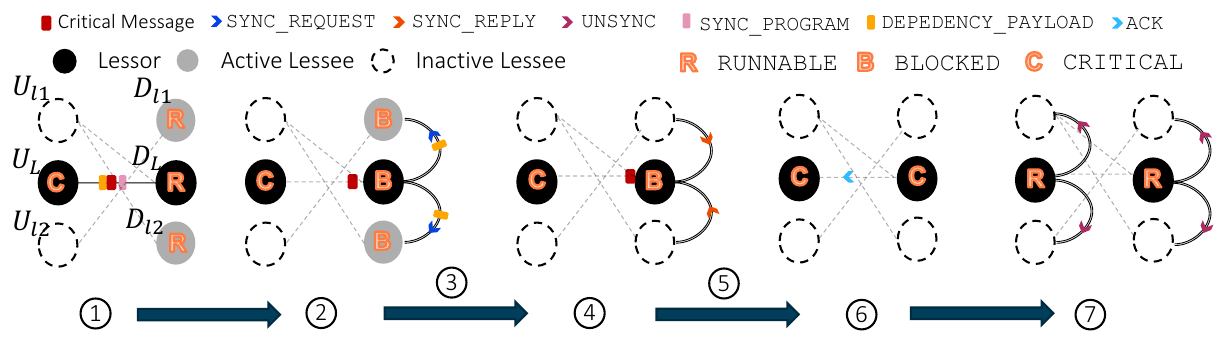}
  }
  \hfill
  \subfloat[\textit{State Transition Timelines for lessor instance $D_{L}$ and lessee $D_{l1}$.
  \label{fig:state-timeline}}]{%
    \includegraphics[width=0.39\textwidth]{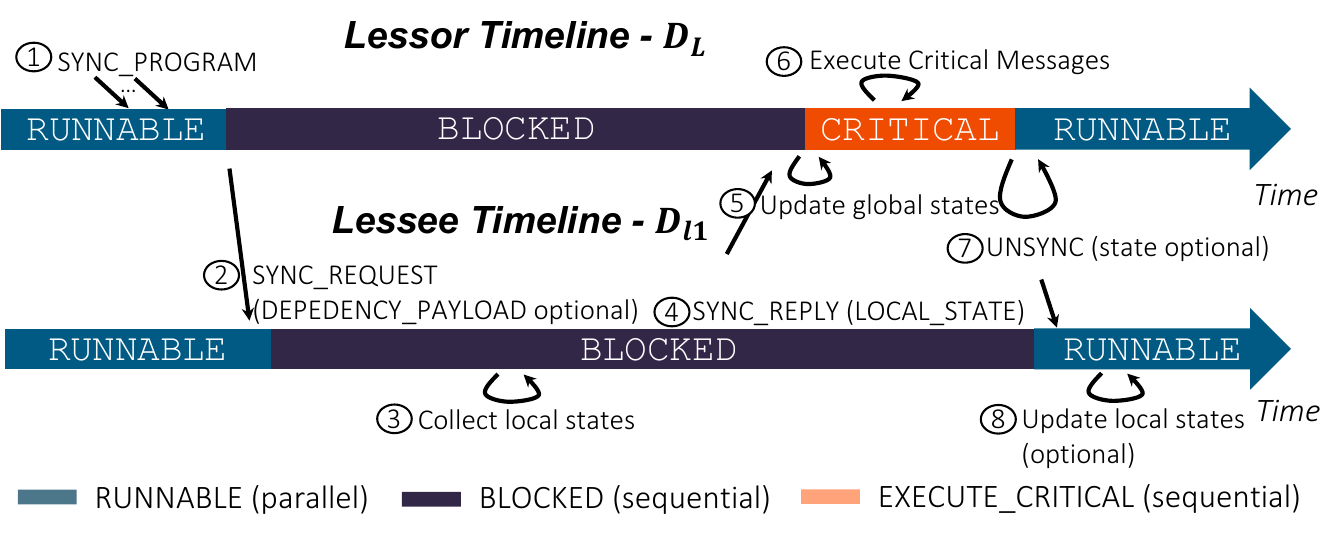}
  }
  \vspace{-.1in}
  \caption{\textit{\small Dual mode actor (\DMA{}) protocol.} 
  \vspace{-.2in}
  }
  \label{fig:2MA-protocol}
\end{figure*}


\subsubsection{\DMA{} Mailbox States}
As described in Section~\ref{sec:overview} each actor has a mailbox that holds all incoming messages, including user and control messages (described below). The \dma{} protocol operates as a state machine over the mailbox state of an actor.
Each mailbox state imposes different restrictions on which messages should be executed or which messages must be blocked. Hence, mailbox states translate into barriers imposed on the execution order of the messages.
In \sysname{}, an actor's mailbox can be associated with one of the three runtime states:


\noindent \textbf{1. \RUNNABLE{} state:}
By default, each mailbox is in the \RUNNABLE{} state, wherein the mailbox will accept any data message to be processed. 
In \RUNNABLE{} state, the mailbox does not restrict the execution order of the messages, implying that messages routed to the lessor instance can be parallelized by routing to lessee instances as well.
This has an interesting nuance: the function state now gets partitioned over the lessor and lessee instances.
We denote these states as \textit{partial states} that is needed to be collected and consolidated when required by the processing semantics of the application (e.g., when a critical message is received).

\noindent \textbf{2. \BLOCKED{} state:}
When a critical message $CM$ arrives at a function instance, 
the mailbox waits for the blocking condition (Section~\ref{subsec:sync-program}) to be satisfied and then switches to the \BLOCKED{} state.
After that, the mailbox blocks all pending set messages $P_{CM}$ and places them in a blocked queue.
In brief, it ensures that all messages in depending set $D_{CM}$ are processed and all messages in pending set $P_{CM}$ 
are blocked.
Once in the \BLOCKED{} state, all parallel instances of the actor can coordinate to consolidate the actor states. This coordination requires control messages to be exchanged between the instances, which we describe below in Section~\ref{subsec:protocol}.

\noindent \textbf{3. \CRITICAL{} state:}
Once the lessor has consolidated the partial states from all instances, the mailbox turns into \CRITICAL{} state, denoting that the instance can now correctly process the critical messages.
The \CRITICAL{} state corresponds to the \SEQ{} mode of execution for the actor. Further, we enforce that in the \SEQ{} mode, all execution happens only at the lessor instance. Hence, only the lessor instance goes in \CRITICAL{} state.
The execution of critical messages marks the completion of the barrier, after which the mailbox can return to the default \RUNNABLE{} state, and the mailbox can then process blocked messages.

\subsubsection{Dual Mode Actor Protocol}
\label{subsec:protocol}
Given the mailbox states and their characteristics, we can now describe the \DMA{} protocol.
We demonstrate the protocol between an upstream actor $U$ (lessor instance $U_{L}$ and parallel lessee instances $U_{l1}$ and $U_{l2}$) and downstream actor $D$ (lessor instance $D_{L}$ and parallel lessee instances $D_{l1}$ and $D_{l2}$) in Figure~\ref{fig:barrier-protocol}. Figure~\ref{fig:state-timeline} shows the corresponding state transition timeline between $D_{L}$ and $D_{l1}$. 

To trigger the \DMA{} protocol, 
$U_{L}$ sends a \textit{SYNC Program} (\SP{}) message along with the critical message (Step \circled{1}) to the lessor instance $D_{L}$.
$U_{L}$ encapsulates essential information about the barrier semantics needed by the associated critical message in the \SP{} message (more details in Section~\ref{subsec:sync-program}).
Once the \SP{} is received, the lessor starts buffering incoming messages from the function that sent the \SP{} message.
The \SP{} message carries along with it: (i) Critical message(s) (that require \SEQ{} mode of execution), (ii) the list of which upstream functions are needed to be blocked for the barrier, and (iii) \texttt{DEPENDENCY\_PAYLOAD} that contains sequence ID of the \textit{last message} sent through \textit{every} active channel between upstream and downstream instances (i.e., $U_{l1}$-$D_{L}$, $U_{l1}$-$D_{l2}$, $U_{L}-D_{L}$, $U_{l2}$-$D_{l1}$) before the barrier was formed on the upstream function $U$.
The lessor instance switches to \BLOCKED{} state after processing all messages that satisfy the blocking condition (details in Section~\ref{subsec:sync-program}).
It then initiates a synchronization process by sending \texttt{SYNC\_REQUEST} messages (step \circled{2}) to all its lessee instances. Additionally, the lessor segregates the sequence IDs received in \texttt{DEPENDENCY\_PAYLOAD} for each lessee channel and sends it as a \texttt{DEPENDENCY\_PAYLOAD} field in \texttt{SYNC\_REQUEST}.
The lessee instances acknowledge receipt of the \texttt{SYNC\_REQUEST} message (this is crucial to ensure that only messages sent before the CM form the barrier - more details in Appendix~\ref{appendix:2MA}).

Upon receiving the \texttt{SYNC\_REQUEST} message, the lessee $D_{l1}$ starts buffering incoming messages while waiting for all messages in the dependency set to be completed (Step \circled{3}).
For $D_{l1}$ with upstream channel $U_{X}-D_{l1} (X=L, l1, l2)$, all messages with a Sequence ID smaller than the sequence ID it received in the \texttt{DEPENDENCY\_PAYLOAD} field, are in the dependency set.
Once the lessee has completed the execution of all messages in the barrier, it also switches to \BLOCKED{} state.
Thereafter, the lessee replies with a \texttt{SYNC\_REPLY} message, which includes any computed partial states and the sequence ID of the latest message that it sent on each downstream channel (Step \circled{4}). 

Once the lessor $D_{L}$ has received \texttt{SYNC\_REPLY} messages from all its lessees, it consolidates the received partial states (Step \circled{5}) and switches to \CRITICAL{} state.
\sysname{} provides a user API with the semantics on how to consolidate the partial states(Section~\ref{subsec:user-api}).
Then $D_{L}$ executes all critical messages (Step \circled{6}) before sending acknowledgment messages (ACKs) to $U_{L}$. Note that the barrier semantics imposed by \SP{} and the consolidation of states (Steps \circled{2}-\circled{4}) ensure that all messages in the dependency set of the critical message have been processed and their outputs are all aggregated at the lessor (details in Section~\ref{subsec:sync-program}.
Once the lessor has processed all critical messages, the barrier is complete at $D$ as well.
Now, if the execution of critical messages at $D$ causes more critical messages to be produced for the further downstream actor, $D_{L}$ will send a \SP{} message (just like the one $U_{L}$ sent at the start of this protocol) and wait for its ACK.
Otherwise, $D$ can directly return to \PARALLEL{} mode of execution.
To do so, the lessor $D_{L}$ unblocks all \BLOCKED{} lessee instances
by sending \texttt{UNSYNC} messages (Step \circled{7}). The lessees turn back to \RUNNABLE{} on receiving the \texttt{UNSYNC} messages.

Note that since critical messages require \SEQ{} execution semantics, the upstream actor $U$ must also be in a \SEQ{} state to ensure that the critical messages are all sequentially sent from the upstream lessor without interleaving any other messages.
This implies that the upstream lessor $U_{L}$ must be in a \CRITICAL{} state, and the upstream lessee instances ($U_{l1}$ and $U_{l2}$) are in a \BLOCKED{} state. For a \SP{} that triggers downstream \SP{}s (e.g., barrier propagation), \texttt{SYNC\_REPLY} serves the dual purpose of carrying partial states (for executing \SP{} within the current actor) and dependency message ID used if new \SP{} is propagated downstream.
All \SP{}s require ACKs sent back to $U_{L}$ before the \texttt{UNSYNC} procedure. 

\noindent\textbf{Lessee Management:}
The \texttt{SYNC\_REQUEST} effectively (i) terminates the lease between the lessor and the lessee instances and (ii) deactivates every channel between upstream and downstream instances. 
To re-activate a channel (say $U_{l1}-D_{l1}$), the upstream instance needs to send a \texttt{LESSEE\_REGISTRATION} message to lessor instance $D_{L}$ (and wait for an acknowledgment) before dispatching a message to $D_{l1}$ (messages are buffered before the reply is received). 
For $D$ that is in the \BLOCKED{} or \CRITICAL{} status, \texttt{LESSEE\_REGISTRATION} will be blocked until $D$ is set back to \RUNNABLE{} state (and therefore $D$ does not create new lessee).
This protocol effectively prevents messages outside of the barrier from being processed on any downstream instances (proof in Appendix~\ref{appendix:2MA}).
Note that a strategy could create lessee instances by forwarding messages from the lessor instance, in which case the lessee identities will be added to the lessor directly without requiring a \texttt{LESSEE\_REGISTRATION} message. 

\subsection{\texttt{SYNC} Program}
\label{subsec:sync-program}

\texttt{SYNC} Program \SP{} is created by scheduling strategy and sent as a message. We list the parameter to an \SP{} in Table~\ref{tbl:sync-params}.
\begin{table}[h!]
\centering
\begin{tabular}{||c | c ||} 
 \hline
\texttt{SYNC} Granularity & \texttt{SYNC\_ONE}, \texttt{SYNC\_CHANNEL} \\ 
\texttt{DEPENDENCY\_PAYLOAD} & \makecell{$\{ s_{kl} \mid$ Sequence ID of the\\ last message from $U_{k}-D_{l}\}$} \\[1ex] 

 \hline
\end{tabular}
\caption{\textit{\small Parameters of \texttt{SYNC} programs.}}
\vspace{-.1in}
\label{tbl:sync-params}
\end{table}



\texttt{SYNC} granularity provides tunable degrees of constraints on which messages to incorporate into the barrier. Consider that a downstream actor $D$ has $N$ upstream actors $U^{i}, 0<i<N$, each mapped to a streaming operator in user-defined DAG. Each upstream actor $U^{i}$ is further parallelized into $P^{i}$ instances $\{U^{i}_{j} \mid 0 < j < P^{i}\}$ ($U^{i}_{0}$ being the lessor instance of actor $U^{i}$). 
We specify the supported \texttt{SYNC} granularity and the blocking conditions as follows:

An upstream instance $U^{i}_{j}$ produces a stream of messages $m(s, U^{i}_{j})$ where $s$ is a monotonically increasing sequence number.
The instance ensures $m(s, U^{i}_{j})$ happens before $m(s', U^{i}_{j})$ ($m(s, U^{i}_{j}) \rightarrow m(s', U^{i}_{j})$) $\forall s < s'$.
Note that critical messages and other control messages used in our protocol above can also be denoted as $m(s, U^{i}_{j})$. For ease of usage, we represent a control message $\mathcal{M}^{i}_{j}$ to denote that the control message $\mathcal{M}$ was sent by upstream instance $U^{i}_{j}$.

\noindent(i)~\textbf{\texttt{SYNC\_CHANNEL}} blocks exactly \textit{one} actor $U^{i}$ (including its instances $\{U^{i}_{j} \mid 0 < j < P^{i}\}$) that sends the \SP{} and also completes all messages sent by the $\{U^{i}_{j} \mid 0 < j < P^{i}\}$ before it switched to \BLOCKED{}.
This allows for the creation of separate barriers for each upstream actor. \texttt{SYNC\_CHANNEL} corresponds to a channel-wise processing barrier (Section~\ref{sec:overview}) and can be particularly useful when propagating watermarks on a single channel, such as windowed aggregation.

\noindent For a \texttt{SYNC\_CHANNEL} barrier triggered by a critical message $CM$ from an upstream actor $U$, we define the \textit{dependency set} of the barrier as follows:

\begin{equation*}
  \begin{split}
    \mathcal{D}_{B_{channel}} = \{ m(s_{j}, U_{j}) \mid &m(s_{0}, U_{0}) \rightarrow CM\ and\ \\ &m(s_{k}, U_{k}) \rightarrow \texttt{SYNC\_REPLY}_{k}, k \neq 0 \}
  \end{split}
\end{equation*}

\noindent The dependency set of the barrier includes the messages that must be processed before the $CM$ of the associated barrier is processed.
The \textit{pending set} of $B$ is the set of messages that must be processed after the $CM$ has been processed. For \texttt{SYNC\_CHANNEL}, we define the pending set as:

\begin{equation*}
  \begin{split}
    \mathcal{P}_{B_{channel}} = \{ m(s_{j}, U_{j}) \mid &CM \rightarrow m(s_{0}, U_{0})\ and\ \\ & \texttt{SYNC\_REPLY}_{k} \rightarrow m(s_{k}, U_{k}), k \neq 0 \}
  \end{split}
\end{equation*}

\noindent (ii)~\textbf{\texttt{SYNC\_ONE}} blocks \textit{all upstream actors} $\{U^{i} \mid 0 < i < N\}$ (including all upstream instances $\bigcup_{i=1}^{N}\{U^{i}_{j} \mid 0 < j < P^{i}\}$) and also completes all running messages sent by $\bigcup_{i=1}^{N}\{U^{i}_{j} \mid 0 < j < P^{i}\}$ before the target downstream actor switches to \BLOCKED{}.
Each upstream actor $U^{i}$ sends a \texttt{SYNC} program from its lessor instance. This allows for creating a barrier that applies to all upstream functions.
\texttt{SYNC\_ONE} corresponds to the global processing barrier (Section~\ref{sec:overview}) - that needs to synchronize across \textit{all} upstream actors.
Scheduling policies can also chain \texttt{SYNC\_ONE} between each pair of upstream/downstream actor to implement distributed snapshot (e.g., checkpoint~\cite{chandy1985distributed}, reconfiguration~\cite{mai2018chi}, etc.)

\noindent If a \texttt{SYNC\_ONE} barrier was formed by critical messages $CM^{i}$ from upstream actor $U^{i}, 0 \leq i \leq N$, we define the \textit{dependency set} of the barrier as follows:

\begin{equation*}
  \begin{split}
    \mathcal{D}_{B_{one}} = \bigcup_{i=1}^{N}D_{CM^{i}} = \bigcup_{i=1}^{N} \{ &m(s^{i}_{j}, U^{i}_{j}) \mid m(s^{i}_{0}, U^{i}_{0}) \rightarrow CM^{i}\ and\ \\ &m(s^{i}_{k}, U^{i}_{k}) \rightarrow \texttt{SYNC\_REPLY}^{i}_{k}, k \neq 0 \}
  \end{split}
\end{equation*}

\noindent The dependency set of the barrier includes the messages that must be processed before the $CM$ of the associated barrier is processed.
The \textit{pending set} of this barrier can be defined as:

\begin{equation*}
  \begin{split}
    \mathcal{P}_{B_{one}} = \bigcup_{i=1}^{N} \{ m(s^{i}_{j}, U^{i}_{j}) \mid &CM^{i} \rightarrow m(s^{i}_{0}, U^{i}_{0})\ and\ \\ &\texttt{SYNC\_REPLY}^{i}_{k} \rightarrow m(s^{i}_{k}, U^{i}_{k}), k \neq 0 \}
  \end{split}
\end{equation*}

Essentially \DMA{} ensures a barrier $B$ is enforced on target actor $D$ (i.e., all parallel instances of $D$) by blocking all messages that belong to the pending set $\mathcal{P}_{B}$ and completing all messages in the dependency set $\mathcal{D}_{B}$. 
For brevity, we formalize and show how \sysname{}'s \DMA{} helps separate dependency set and pending set in Appendix~\ref{appendix:2MA}.

%% file: text/5-design.tex
\section{\sysname{} Scheduling}
\label{sec:scheduling}
In this section we discuss internal mechanism of \sysname{} runtime and how \sysname{} uses customized scheduling strategies to improve performance isolation while sharing resources. 

\subsection{Scheduling Strategy API}





In \sysname{}, all messages that are not blocked by the \DMA{} protocol, are marked for scheduling. A per-worker scheduling strategy takes in all such messages from all functions and performs the following tasks: (i) schedules the highest priority message to be executed by the worker, and (ii) routes messages to lessee instances in the face of SLO violations.
Table~\ref{tab:api} shows \sysname{}'s scheduling API along with the hooks at which the exposed API runs and the description of a use case for the hook.

For a message that arrives at an actor mailbox (whether lessor or lessee instance), the fetcher thread first invokes \texttt{enqueue} function, which allows the strategy to make forwarding decisions on the message.
Once the message passes the \texttt{enqueue} function, it gets added to the actor mailbox and is available for scheduling on the worker.
The \sysname{} worker thread, in turn, continuously calls \texttt{getNextMessage} to get the next scheduled message from the strategy.
Note that in \sysname{}, we adopt a multiplexed resource design, where the strategy can schedule messages across jobs and functions. 
Once a message is scheduled, \sysname{} calls \texttt{preApply} and \texttt{postApply} before and after the message is executed on the worker thread. Finally, a message's execution can result in another message targetting another function. To modify this newly created message, \sysname{} exposes a \texttt{prepareSend} hook.
This hook is also used to dispatch \SP{} message (critical for the \DMA{} protocol) when a critical message is produced at a function. 

The scheduling strategy could maintain its own data structures to schedule messages (e.g., based on priority, arrival order, etc.) and choose parallel lessees (based on load balancing strategies).



\begin{table}
  \begin{minipage}{\columnwidth}
      \centering
      \begin{scriptsize}
      \begin{tabular}{| m{7.5em} | m{10em} | m{12em} |} 
          \hline
          {\bf API} & {\bf Hook} & {\bf Usecase} \\ 
          \hline
          {\texttt{enqueue()}} & Called by fetcher thread upon receiving message. & Whether the message should be executed locally or it should be forwarded to a lessee. \\ 
          \hline
          {\texttt{getNextMessage()}} & Called by worker to get next message to process. & Strategy chooses the message with highest priority \textit{across all functions} on the worker. \\
          \hline
          {\texttt{preApply()}} & Called by worker before executing function on the chosen message. & Collect per-message execution time by starting a timer. \\
          \hline
          {\texttt{prepareSend()}} & Called by worker before sending an output message from the function. & Change the recipient address of the function from a user address to a downstream lessee instance. \\
          \hline
          {\texttt{postApply()}} & Called by worker right after executing the function. & Collect per-message execution time. \\
          \hline
      \end{tabular}
      \end{scriptsize}
      \caption{\textit{\small Scheduling API and hooks in \sysname{}}}
      \label{tab:api}
      \vspace{-.1in}
  \end{minipage}
\end{table}

\subsection{\sysname{} Scheduling Strategies}


As described in Section~\ref{sec:overview}, in \sysname{} requests are forwarded in the form of messages. Hence, autoscaling actors by offloading messages also require messages to be sent to the appropriate lessee. In \sysname{}, we support two modes of autoscaling, based on where the autoscaling happens:
\newline\noindent (i) \textbf{A \RSSHARED{} approach}, a \textit{lessor initiated} approach where all messages from upstream are sent to the downstream lessor instance and the scheduling policy at the worker running the lessor, decides whether and where to forward the message (through \texttt{enqueue} function API).

\noindent (ii) \textbf{A \DSSHARED{} approach}, an \textit{upstream initiated} approach that directly modifies the address on the message using the \texttt{prepareSend} API to dispatch the message to a lessee instance (already registered using the \texttt{LESSEE\_REGISTRATION} message). 

\begin{figure}[!tp]
    \centering
    \includegraphics[width=.48\textwidth]{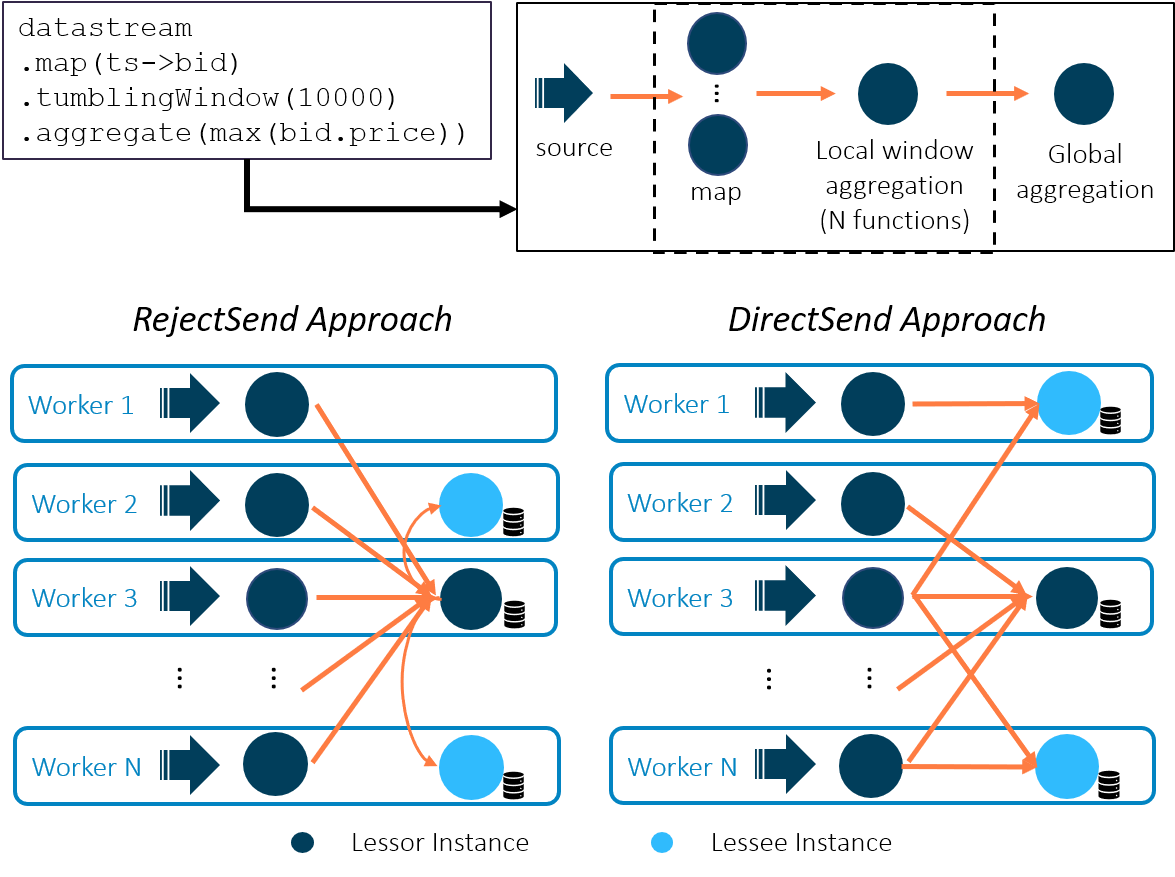}
    \caption{\textit{\small Message path of \RSSHARED{} vs. \DSSHARED{}.}}
    \vspace{-.2in}
    \label{fig:scheduling-scenarios}
\end{figure}

We now discuss the utility of the two modes of autoscaling.
Figure~\ref{fig:scheduling-scenarios} depicts a streaming dataflow deployed in a cluster of $N$ workers. The query given on the left is implemented by the user in the form of the logical plan shown on the right. 
The dataflow DAG contains a set of stage 1 map operators (each mapped to a data source), stage 2 local window aggregate operators (each performing max over the local windows) and a final stage 3 global aggregate operator that computes the \texttt{max} aggregation of windows from stage 2 operators. 
The user divides the \texttt{max} operation into two stages so that the first stage can have multiple functions running concurrently and a final global aggregate computes the global max over all values returned from the stage 2 functions.
In this dataflow, even after having multiple functions for local aggregation, the stage 2 operator could become a bottleneck operator as the input rate from stage 1 increases.

We compare a \RSSHARED{} strategy against a \DSSHARED{} strategy by parallelizing \textit{each} function of the stage 2 operator in the two modes as shown in Figure~\ref{fig:scheduling-scenarios}.
We deploy the shown logical plan on a cluster of 128 workers. Each stage-1 operator is run as one function per worker. Each stage-2 operator is run as $n$ lessor functions, with each lessor having a maximum of $m$ lessee functions. We run our experiments for four configurations of $(n, m) = \{(64, 2), (32, 4), (16, 8), (8, 16)\}$.
Finally, a single instance of stage-3 operator is run. We then run experiments to evaluate the two strategies for the following two aspects:

\begin{enumerate}[noitemsep,nolistsep]
  \item \textit{Load Balancing}: We run the two strategies to randomly choose stage-2 lessees. We compare which strategy gives a better performance as we scale the number of lessee instances for the stage-2 function.
  \item \textit{SLO Satisfaction}: We run the two strategies such that messages are routed to the lessees when the lessor is guaranteed to have a SLO violation on the message. We run this experiment for varying skewness of the workload (one stage-1 function is generating more messages than another).
\end{enumerate}




\begin{figure}[!tp]
    \subfloat[\textit{Job latencies distribution with increasing number of number of parallel instances per operator.~\label{fig:direct-forward-span}}]{%
      \includegraphics[width=0.23\textwidth]{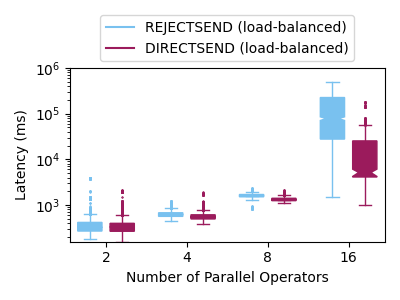}
    }
    \hfill
    \subfloat[\textit{Job latencies distribution with increasing load skewness with zipfian distribution input rate.\label{fig:direct-forward-skew}}]{%
      \includegraphics[width=0.23\textwidth]{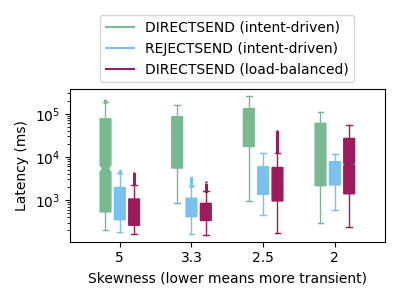}
    }
    \caption{\textit{\small Comparison of two scheduling approaches.} 
    }
    \vspace{-.1in}
    \label{fig:direct-forward-comparison}
\end{figure}

\noindent \textbf{\DSSHARED{} scales better than \RSSHARED{} policies.} 
Figure~\ref{fig:direct-forward-span} shows that \DSSHARED{} provides better latencies benefit compared to \RSSHARED{}, especially when the number of parallel instances for each function is high.
This is because \RSSHARED{} forwards messages from the lessor, so even though the lessor does not process the message, the message is needed to be deserialized, run by the strategy and then forwarded.
When the number of lessees is large, the synchronization phase in \DMA{} protocol runs longer and hence, this per-message overhead at the lessor worsens.
On the other hand, \DSSHARED{} evenly distribute messages from upstream instance directly, distributing the overhead of message parsing and forwarding.

\noindent\textbf{\RSSHARED{} responds to workload skewness better than \DSSHARED{}.} Figure~\ref{fig:direct-forward-skew} shows that SLO-driven \RSSHARED{} performs better than SLO-driven \DSSHARED{} with 3.2$\times$ median latency reduction for skewed workloads.
This is because in SLO-driven \DSSHARED{} strategy, the upstream instances pauses messages to any downstream instance that had a SLO violation for a fixed period of time. In a sense, the upstream instances get delayed information about violations at the downstream while \RSSHARED{} 
performs message forwarding directly on the instance of SLO violation avoiding delays.
Note that the delayed information in SLO-driven \DSSHARED{} leads to worse performance even compared to the SLO-unaware \DSSHARED{} strategy (with up to 2$\times$ median latency increase).  

\subsection{User API}
\label{subsec:user-api}
\sysname{} exposes API for users to describe the type of state being computed by a function, its processing semantics and how to consolidate partial states accumulated on multiple instances of the same function.
\sysname{} provides the following managed data structures:
(i) $\afunc{ValueState\langle T \rangle}$
, (ii) $\afunc{ListState\langle T \rangle}$
, and (iii) $\afunc{MapState \langle K, V \rangle}$
.
By default, \sysname{} assumes all operators are parallelizable (and has \RUNNABLE{} state by default). \sysname{} scheduling strategy is responsible for identifying associative-decomposable operations~\cite{yu2009distributed}.
Most messages consumed by \textit{stateless} streaming operators fall under this category (e.g., $\afunc{map}$, $\afunc{filter}$, $\afunc{keyBy}$, etc.).
For stateful operators (e.g., $\afunc{windowApply}$, $\afunc{windowReduce}$, $\afunc{join}$, etc.), \sysname{} lets users specify a \texttt{CombiningFunction} $f(T, T)\xrightarrow{}T$ in order to allow partial states to be combined during the \DMA{} procedure.
For the common-case stateful operations with bounded state size (distributive and algebraic aggregation~\cite{gray1997data}) such as $\afunc{sum}$, $\afunc{max}$, $\afunc{average}$, etc. 
each lessee instance performs calculates partial states before sending to the lessor where the states get aggregated as per the \texttt{CombiningFunction}.
For the other stateful operations with unbounded state size (holistic aggregation~\cite{gray1997data}) such as $\afunc{median}$ and $\afunc{histogram}$, partial states are a $ListState<T>$ of all updates. These can be appended together before the \texttt{CombiningFunction} is applied.

\section{Implementation}
\label{sec:implementation}

\sysname{} is built on top of Flink Statefun~\cite{statefun}, which uses Flink as an underlying message processor. \sysname{}'s runtime manages all actor instances and actor states. The runtime enqueues (through fetcher thread) and processes (through worker thread) user messages (Section~\ref{sec:overview}) and performs \DMA{} protocol in Section~\ref{sec:dual-mode-actor}.
\sysname{} exposes the Scheduling API as mentioned in Section~\ref{sec:scheduling}.

\noindent We modify the Flink Statefun runtime to support custom scheduling strategies and implement the \DMA{} protocol along with the control messages required for the protocol. Few important aspects of the implementation are highlighted below:

\noindent\textbf{Read-heavy workload:}
The \DMA{} protocol causes synchronization for every state-read critical message. A small tweak in the \DMA{} protocol can make \sysname{} suitable for read-heavy workloads as well.
The \texttt{UNSYNC} message can carry back the consolidated state and each read operation can be performed on the lessees independently.

\noindent\textbf{Message Overheads:}
The \SP{} control message and the $CM$ from the upstream lessor are sent between the same upstream lessor and downstream lessor. Hence, to avoid message serialization and deserialization costs, we merge the \SP{} and $CM$s into a single message.

%% file: text/6-eval.tex
\section{Evaluation}
\label{sec:evaluation}
In this section, we evaluate \sysname{} by investigating the following questions: (i)~\textit{How does \sysname{} help DSPS adapt to transient load changes?} (ii)~\textit{What is the overhead of the \DMA{} protocol?} and (iii)~\textit{Could \sysname{} support other SLO beyond latency?}

By default, we use a 32-node cluster of xl170 Cloudlab machines~\cite{Duplyakin+:ATC19} (10 2.4GHz cores/64 GB memory/10GB Bandwidth). Our default setting runs \sysname{} using 128 workers. We adopt queries from Nexmark~\cite{nexmark} and implement dataflows of functions corresponding to queries from the benchmark set (Q7, Q12). To determine latency SLO, we run a single dataflow job in the default setting with a constant input volume so that the cluster utilization reaches approximately 50\%. We set the latency SLO to be twice the tail latency we collect in this setting. 


\begin{figure}[!htp]
    \centering
    \includegraphics[width=.5\textwidth]{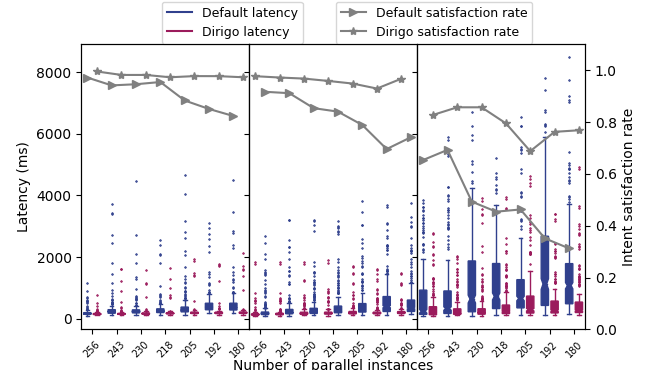}
    \caption{\textit{\small \sysname{} vs. Default scheduling strategy under increasing workload transiency (Pareto distribution with decreasing $\alpha$). With $\alpha$ set to 5, 3.3 and 2.5. Boxes show latency distribution. Scatter dots show SLO satisfaction rate of each setting.}}
    \label{fig:varying-resource-zipf}
    \vspace{-.1in}
\end{figure}

\noindent\textbf{\sysname{} utilizes resources better to satisfy SLO under transient load changes.} Figure~\ref{fig:varying-resource-zipf} shows \sysname{}'s benefit (with \RSSHARED{} policy) over the default scheduling strategy (FIFO without auto-scaling) while using \textit{two} running jobs (with initial dataflow parallelism of 128) in a cluster. We fix the number of jobs and decrease the number of \sysname{} workers provisioned by \sysname{} cluster. 
When using the maximum number of workers (256), the two dataflows map to actors with lessor instances placed on machines that do not overlap (machine 1-16 for job 1 and machine 17-32 for job 2). 
We apply workload changes that follow Pareto distribution with increasing skewness ($\alpha = 5, 3.3, 2.5$).
Our results show that 
(i) In all our scenarios, \sysname{} can provide an equal or better overall satisfaction rate (equal, 5\%, 12\% increase respectively) with 30\% resource savings than using the default strategy in an isolated setting. 
(ii) Na\"ively collocate functions lead to satisfaction rate degradation ranging from 15-34\% (with median latency increase up to 4.5$\times$) when reducing resource consumption by 30\%. \sysname{} scheduling policy improves performance isolation by controlling satisfaction rate drop within 1-14\% and (maximum median latency increase by 1.8$\times$). 
(iii) Workload transiency widens the benefits of using the \sysname{} strategy as \sysname{} provides increasing satisfaction rate improvement (up to 15, 23, 46\%) and tail latency (99p) as we increase the workload transiency. It also provides latency reduction up to 2$\times$, 2.7$\times$, and 1.8$\times$, respectively. These results show that \sysname{}'s data-plane scheduling could react to workload changes quickly.

\begin{figure}[!htp]
  \vspace{-.1in}
    \subfloat[\textit{\DMA{} overhead over increasing number of parallel lessee instances (state size 1K)~\label{fig:sync-overhead-parallelism}}]{%
      \includegraphics[width=0.22\textwidth]{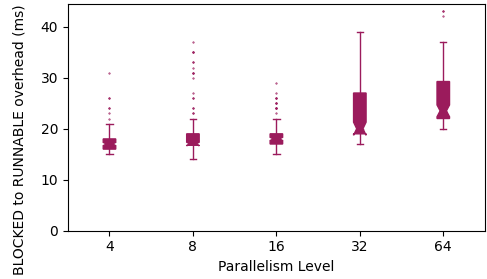}
    }
    \hfill
    \subfloat[\textit{\DMA{} overhead over increasing state sizes collected (parallelism level 4)~\label{fig:sync-overhead-statesize}}]{%
      \includegraphics[width=0.22\textwidth]{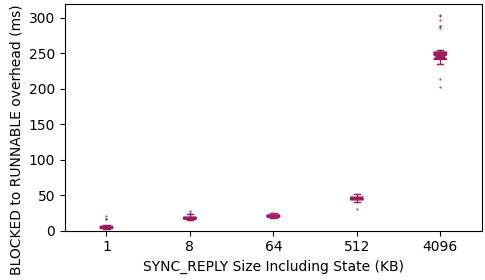}
    }
    \caption{\textit{\small \DMA{} protocol overhead.} 
    }
    \vspace{-.1in}
    \label{fig:sync-overhead}
\end{figure}

\noindent\textbf{\DMA{} scales with the number of parallel instances and state sizes.} Figure~\ref{fig:sync-overhead} shows the scalability of the \DMA{} protocol by evaluating the overhead of \DMA{} over an increasing number of parallel instances and state sizes collected using \texttt{SYNC\_REPLY}. We measure the overhead of \DMA{} by recording the duration between the lessor instance turning to \BLOCKED{} state until the last lessee instance receives \texttt{UNSYNC} message. 
Figure~\ref{fig:sync-overhead-parallelism} shows that \DMA{} overhead stays relatively unchanged up to 16 instances (lessees). A higher level of scaling factor (32 and 64) lessees introduces higher synchronization overhead (median latency increases by 18\% and 41\%). This result indicates that for applications that have a lax performance target (> 300-400ms), \DMA{} provides the benefits of autoscaling without resulting in a significant burden on the performance (less than 10\%), despite the input pattern. For an application that is tightly latency-constraint, it is recommended to use a smaller parallelism level (less than 32) and perform logical plan change (e.g., through reconfiguration) when needed. 
The overhead of \DMA{} remains stable (below 20ms) as we increase the partial state size until the size of 64KB (Figure~\ref{fig:sync-overhead-statesize}). 
Transporting state objects of sizes 512KB and 4096KB has a noticeable impact on the overhead (2.3$\times$ and 12.3$\times$, respectively). 
While most of the queries in our experiments use small states ($<$1K), for queries that do generate large function state, it is recommended that the scheduling strategy should (i) deprioritizing auto-scaling functions that generate large state sizes and (ii) uses lower parallelization level or perform user-level state partition.

\begin{figure}[htp]
    \centering
    \includegraphics[width=.45\textwidth]{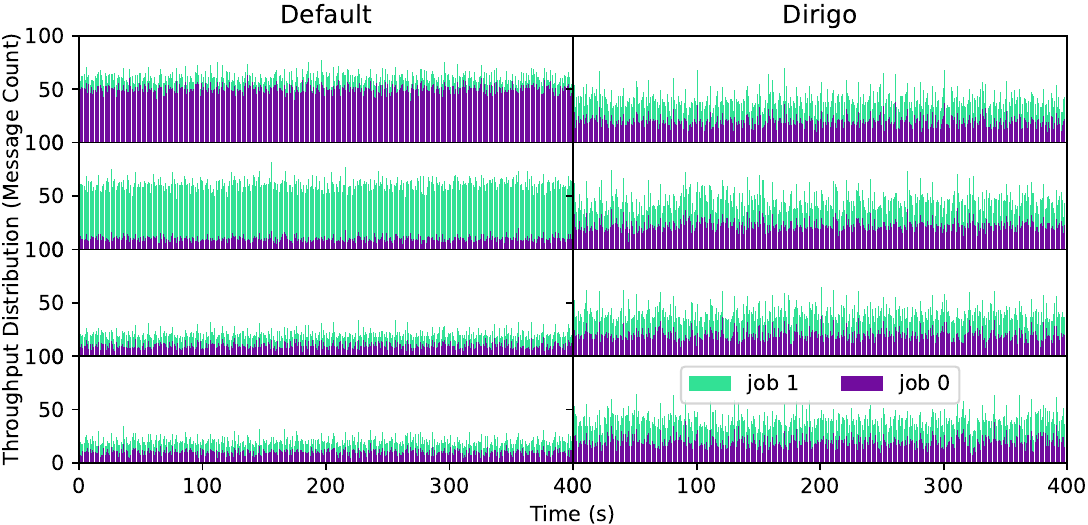}
    \caption{\textit{\small \sysname{} performance load-balancing based using rate-control mechanism.}}
    \vspace{-.1in}
    \label{fig:rate-control}
\end{figure}

\noindent\textbf{\sysname{}'s scheduling API supports SLO beyond latency requirements.} While most scenarios we discuss are driven by latency-target, we show that \sysname{} supports other types of SLO through scheduling API. Figure~\ref{fig:rate-control} shows Q12 with an imbalanced distribution due to the skewed key distribution of user IDs. In this experiment, we perform load-balancing by assigning an equal amount of tokens to messages targeting each job on each worker. The input message that does not receive tokens is processed at a lowered priority and scattered to the remaining workers. The results show that through the use of scheduling API, \sysname{} is not only able to achieve throughput isolation at the worker level but also helps messages to be evenly distributed throughout all workers. 

%% file: text/7-related.tex
\vspace{-2mm}
\section{Related Works}
\label{sec:related}
\vspace{-2mm}



\sysname{}'s architecture supports fine-grained auto-scaling. We categorize existing techniques as the following:

\noindent{\textbf{Elasticity for real-time dataflow applications:}}
State-of-the-art cloud based DSPS largely assume slot-based resource provisioning (e.g. resource containers).
These approaches use mechanisms that constantly allocate and de-allocate resources in response to load changes through reconfiguration~\cite{castro2013integrating, floratou2017dhalion,garefalakis2018medea,li2015supporting, heinze2015online, kalavri2018three,xu2016stela, hoffmann2019megaphone, kalavri2020support}. Reconfiguration approaches are complementary to resource provisioning in \sysname{} --- they allow more complex modification to the dataflow (e.g., operator fusion and operator replacement) at a coarser time granularity (typically from the minimum of several seconds~\cite{kalavri2018three} to a maximum of hours~\cite{wang2022non} depend on the dataflow size). Therefore, slot-based architectured DSPS with reconfiguration is unsuitable for sharing resources at message-level granularity. 
To mitigate long-term workload changes, \sysname{}'s \DMA{} supports a reconfiguration mechanism.
Slot-based architectures do not share resources among jobs. Existing multi-tenant provisioning techniques~\cite{henge} have chosen reactive and coarse-grained (10s re-evaluation interval) mechanisms.


On the other hand, traditional DSMS systems~\cite{abadi2003aurorafull, borealis, madden2002continuously, chandrasekaran2003telegraphcq1} adopt an event-driven architecture with scheduling techniques~\cite{carney2003operator, babcock2003chain} that eliminate resource under-utilization~\cite{welsh2001seda}, which is akin to a serverless architecture.  \sysname{} adopts a similar event-driven design but also improves scalability over traditional DSMS by incorporating operator scaling into scheduling decisions.  
Recent works on fine-grained resource provisioning for DSPSes such as Cameo~\cite{cameo} and EdgeWise~\cite{fu2019edgewise} focuses on ordering and scheduling events targeting streaming operators locally within each machine instance, while \sysname{} supports not only event scheduling within each machine, but also dynamic operator scaling by scheduling events across machine boundaries. 

\noindent{\textbf{Provisioning for serverless architectures:}}
Most works focusing on resource provisioning for serverless architecture assume that function scheduling can be performed by \textit{decoupled} \textit{control plane} (semi-) global entities~\cite{singhvi2021atoll, sreekanti2020cloudburst} or client~\cite{tariq2020sequoia, romero2021faa}. \sysname{} on the other hand adopt a data plane approach to perform message-level provisioning with minimum scheduling overhead. Systems such as~\cite{moritz2018ray, carver2020wukong} adopt a similar scheduler architecture by supporting data-plane function scheduling, but neither of these works support SLO driven scheduling policies. 
Decentralized/hybrid fine-grained scheduling mechanism has been explored in general cloud frameworks before serverless computing platforms have emerged\cite{ousterhout2013sparrow, mitzenmacher2001power, dean2013tail, delgado2015hawk, delimitrou2015tarcil, rasley2016efficient, delgado2016job}. While none of these scheduling mechanisms are designed specifically for DSPS, \sysname{} provides an opportunity to explore the effects of these scheduling technqiues within DSPS through its flexible scheduling frameworks and policy API. 


\noindent{\textbf{Scaling fine-grained stateful functions:}}
As previously discussed in Section~\ref{subsec:existing-approaches}, \sysname{} is different from existing stateful serverless frameworks such as Netherite~\cite{burckhardt2021durable, burckhardt2021serverless, burckhardt2022netherite} and Hydro~\cite{sreekanti2020cloudburst, wu2020transactional} by proposing a mechanism that ensures orderliness of critical events that target both stateless and stateful function, without relying on external storage for coordination. 
Similar to \sysname{}, DPA~\cite{kraftdata} proposes an actor-based programming model that enables various modes of parallel query execution. Similar to Netherite~\cite{burckhardt2022netherite}, its parallel operators are pre-determined by query compiler and therefore does not require actor states to be dynamically scaled-out/in as in \sysname{}. 

Dynamic scale-out/in for fine-grained stateful tasks is not unique to dataflow applications.
Network functions (NFs) also proposes \textit{native} state handling techniques during NF scaling~\cite{rajagopalan2013split, gember2014opennf, khalid2019correctness}. 
\sysname{}'s scheduling mechanism supports diverse needs of workloads~\cite{rajagopalan2013split, gember2014opennf, khalid2019correctness} through its scheduling policies. 
\sysname{} assumes a fully-decentralized system architecture, whereas existing NF deployment requires partially centralized entities (controllers and switches) during the coordination process. \sysname{}'s \DMA{} is also similar to 1Pipe~\cite{li20211pipe}, while 1Pipe focuses on failure handling but \sysname{} focuses on synchronization during dynamic scale-out/-in.


%% file: text/8-conclusion.tex
\vspace{-2mm}
\section{Conclusion}
\label{sec:conclusion}
\vspace{-2mm}

This paper proposes \sysname{}, a distributed stream processing service built upon virtual actors.
\sysname{} addresses the dual challenge of resource utilization and performance isolation for stream processing applications by first adopting a serverless paradigm that enables resource sharing. 
Then \sysname{} proposes (i) dual-mode actor protocol that supports ordering requirements of streaming operators during autoscaling, and (ii) data plane scheduling API that helps customized, SLO-driven strategy to improve performance isolation.


%% file: text/9-appendix.tex
\section{\DMA{} Protocol Blocking Conditions}
\label{appendix:2MA}
Given the above two granularities for synchronization, we now specify the blocking conditions required by the two primitives and then show how \sysname{} achieves these conditions by the use of its special messages and the \DMA{} protocol.

Consider that a downstream actor $D$ has $N$ upstream actors $U^{i}, 0<i<N$, each mapped to a streaming operator in user-defined DAG.
In \sysname{}, each upstream actor $U^{i}$ could be further parallelized into $P^{i}$ instances $\{U^{i}_{j} \mid 0 < j < P^{i}\}$ with $U^{i}_{0}$ being the lessor instance of actor $U^{i}$.
We optionally drop $i$ in the notation and use just $U_{j}$ to refer to the $j$ parallel instances of the upstream actor, if there's only one upstream actor of concern. 
An upstream instance $U^{i}_{j}$ produces a stream of messages $m(s, U^{i}_{j})$ where $s$ is a monotonically increasing sequence number.
The instance ensures $m(s, U^{i}_{j})$ happens before $m(s', U^{i}_{j})$ ($m(s, U^{i}_{j}) \rightarrow m(s', U^{i}_{j})$) $\forall s < s'$.
Note that critical messages and other control messages used in our protocol above are special cases of the messages $m(s, U^{i}_{j})$. We denote a control message $\mathcal{M}^{i}_{j}$ to denote that the control message $\mathcal{M}$ was sent by upstream instance $U^{i}_{j}$.
\newline Now consider a critical message $CM^{i}$ being sent from an upstream actor $U^{i}$ to a downstream actor $D$.
The \textit{dependency set} for $CM^{i}$ is the set of messages that were sent from actor $U^{i}$ before $U^{i}$ converted into \SEQ{} state to send out $CM^{i}$:

\begin{equation*}
  \begin{split}
    \mathcal{D}_{CM^{i}} =& \{ m(s^{i}_{j}, U^{i}_{j}) \mid m(s^{i}_{0}, U^{i}_{0}) \rightarrow CM^{i}\ and\ \\ &m(s^{i}_{k}, U^{i}_{k}) \rightarrow \texttt{SYNC\_REPLY}, k \neq 0 \}
  \end{split}
\end{equation*}

The \textit{pending set} for $CM^{i}$ is the set of messages that are sent from actor $U^{i}$ that need to be processed after $CM^{i}$ is fully processed. 
\begin{equation*}
    \begin{split}
      \mathcal{P}_{CM^{i}} =& \{ m(s^{i}_{j}, U^{i}_{j}) \mid CM^{i} \rightarrow m(s^{i}_{0}, U^{i}_{0}) \ and\ \\ &\texttt{SYNC\_REPLY}\rightarrow m(s^{i}_{k}, U^{i}_{k}), k \neq 0 \}
    \end{split}
\end{equation*}

\noindent\textbf{\texttt{SYNC\_CHANNEL} Barrier}: For a \texttt{SYNC\_CHANNEL} barrier, a barrier is needed to be constructed between an upstream actor $U$ and a downstream actor $D$.
In this case, barrier $B_{channel} = \{CM_{j} \mid 0<j<N\}$ between $U \xrightarrow{} D$, where all critical messages $CM_{j}$ must be sent by the same upstream lessor instance $U_{0}$. Hence, the dependency of the barrier $D_{B}$ is equivalent to the dependency set of any one of the critical messages sent by $U_{0}$:
\begin{equation*}
  \begin{split}
    \mathcal{D}_{B_{channel}} = \{ m(s_{j}, U_{j}) \mid &m(s_{0}, U_{0}) \rightarrow CM\ and\ \\ &m(s_{k}, U_{k}) \rightarrow \texttt{SYNC\_REPLY}_{k}, k \neq 0 \}
  \end{split}
\end{equation*}

We will now argue that the \DMA{} protocol between upstream actor $U$ and downstream actor $D$ ensures that all messages $m \in \mathcal{D}_{B_{channel}}$ will be processed before any of the $CM_{j} \in B_{channel}$ is processed.
Now consider the set of messages $m_{D_{k}} \subset \mathcal{D}_{B_{channel}}$ that were processed by downstream instance $D_{k}$. We analyze two cases as follows:
\begin{enumerate}
  \item Let us first consider the downstream lessor instance $D_{0}$.
  \begin{enumerate}
    \item Any messages in $m \in m_{D_{0}}$ that originated from $U_{0}$ are trivially satisfied to be processed before $CM$, because of the happens before relation between $m$ and $CM$ on instance $U_{0}$ and because both $m$ and $CM$ travel on the same $U_{0}-D_{0}$ link. Hence, $m$ is guaranteed to be processed before $CM$.
    \item Any messages in $m \in m_{D_{0}}$ that originated from $U_{k}, k \neq 0$ (an upstream lessee) will satisfy the following relation: let $s_{k}$ be the Sequence ID of the last message sent by $U_{k}$ on the channel $U_{k}-D_{0}$ before it went into a \BLOCKED{} state. Then, $m \rightarrow m(s_{k}, U_{k})$ ($s_{k}$ is the last message to be included on the channel), $m(s_{k}, U_{k}) \rightarrow \texttt{SYNC\_REPLY}_{k}$ and $\texttt{SYNC\_REPLY}_{k} \rightarrow CM$ (since $U_{0}$ will only send critical messages once it has received $\texttt{SYNC\_REPLY}_{k}$ from all its lessees). By transitive property, $m \rightarrow CM$ at $D_{0}$. Hence, all mesages $m \in m_{D_{0}}$ also form the barrier.
  \end{enumerate}
  \item Consider the set of messages $m_{D_{l}}, l\neq 0$, received on a downstream lessee instance.
  For any upstream $U_{k}$, the following relation holds: let $s_{kl}$ be the Sequence ID of the last message sent by $U_{k}$ on the channel $U_{k}-D_{l}$ before it went into a \BLOCKED{} state. Then $m(s_{kl}, U_{k}) \rightarrow \texttt{SYNC\_REPLY}_{k}$ and $\texttt{SYNC\_REPLY}_{k} \rightarrow SP$ (since $U_{0}$ will only send critical messages once it has received $\texttt{SYNC\_REPLY}_{k}$ from all its lessees) and $SP \rightarrow \texttt{SYNC\_REQUEST}_{l}$ (at the downstream lessor $D_{0}$). By transitive property, $m(s_{kl}, U_{k}) \rightarrow \texttt{SYNC\_REQUEST}_{l}$ at the downstream lessee instance $D_{l}$. Note that $D_{l}$ will only block messages after it receives the \texttt{DEPENDENCY\_PAYLOAD} in the $\texttt{SYNC\_REQUEST}_{l}$ message from $D_{0}$. Hence, all mesages $m(s_{kl}, U_{k})$ also form the barrier.
\end{enumerate}

The above proves that all messages in the barrier $B_{channel}$ are processed before the critical message $CM$ is processed.
Next, we also argue that any message that use $B_{channel}$ as dependency (that is, all messages in the \textit{pending set}) shall be blocked and will not be processed until the critical message is processed at downstream lessor $D_{0}$. We define the pending set of a \texttt{SYNC\_CHANNEL} barrier as:

\begin{equation*}
  \begin{split}
    \mathcal{P}_{B_{channel}} = \{ m(s_{j}, U_{j}) \mid &CM \rightarrow m(s_{0}, U_{0})\ and\ \\ & \texttt{SYNC\_REPLY}_{k} \rightarrow m(s_{k}, U_{k}), k \neq 0 \}
  \end{split}
\end{equation*}

First, any message $m \in \mathcal{P}_{B_{channel}}$ such that $CM \rightarrow m$ on $U_{0}$, will be blocked at the upstream lessor since $U_{0}$ will convert into a \BLOCKED{} state before sending the $CM$.
Similarly, for a message $m \in \mathcal{P}_{B_{channel}}$, such that ${SYNC\_REPLY}_{k} \rightarrow m$, $m$ will remain blocked on the upstream lessee.
Once the downstream lessor has processed the \SP{} message, the upstream instances can get unblocked. Note that at this time, downstream instances may not have met the blocking condition (that is, not all messages in the dependency set $D_{B_{channel}}$ might have been processed).
The messages in $P_{B_{channel}}$ on the upstream instances remain blocked until the downstream lessor has received and processed the sync message \SP{}.
The processing of \SP{} implies that the \texttt{DEPENDENCY\_PAYLOAD} must be sent to the downstream lessees (Step \circled{2} in \ref{fig:barrier-protocol}).
This ensures that all downstream instances (including the lessees) know the last Sequence ID (and therefore the last message) sent from an upstream instance.
Consider the Sequence ID $s_{kl}$, received at downstream instance $D_{l}$ in the \texttt{DEPENDENCY\_PAYLOAD} field, for upstream channel $U_{k}-D_{l}$. We consider the following two cases for $D_{l}$:

\begin{enumerate}
  \item $D_{l}$ is downstream lessor ($l = 0$): Upstream instances unblock only after $D_{0}$ sends an acknowledgement for \SP{}, and by processing \SP{}, $D_{0}$ knows the Sequence ID $s_{k0}$ for each upstream instance $U_{k}$.
  \item $D_{l}$ is a lessee instance ($l \neq 0$): The \texttt{SYNC\_REPLY} message acts as a termination of all channels for upstream lessee instances $U_{k}$. For $U_{k}$ to send messages to $D_{l}$, a \texttt{LESSEE\_REGISTRATION} message is needed to be sent to $D_{0}$. Formally, $\texttt{LESSEE\_REGISTRATION}_{k} \rightarrow m(s, U_{k}) \forall s > s_{kl}$. Further, the \texttt{LESSEE\_REGISTRATION} message shall only be accepted after the \SP{} message has been processed at $D_{0}$. Hence, $D_{l}$ is informed of $s_{kl}$ before messages $m \in \mathcal{P}_{B_{channel}}$ are sent by upstream lessees. 
\end{enumerate}

This completes our proof that all messages $m \in \mathcal{D}_{B_{channel}}$ are included in the barrier and all messages $m \in \mathcal{P}_{B_{channel}}$ are blocked out.

\noindent\textbf{\texttt{SYNC\_ONE} Barrier}: For a \texttt{SYNC\_ONE} barrier, $B_{one} = \{CM^{i} \mid 0<i<N\}$ is created between a set of upstream actors $U^{i}$ and a downstream actor $D$. $CM^{i}$ is sent by upstream lessor instance $U^{i}_{0}$, and each $CM^{i}$ has a dependency set $D_{CM^{i}}$.
In this case, each critical message is sent from a different upstream lessor instance and hence, the dependency set for the barrier $D_{B_{one}}$ is the union of the dependencies of each of the critical messages.
\begin{equation*}
  \begin{split}
    \mathcal{D}_{B_{one}} = \bigcup_{i=1}^{N}D_{CM^{i}} = \bigcup_{i=1}^{N} \{ &m(s^{i}_{j}, U^{i}_{j}) \mid m(s^{i}_{0}, U^{i}_{0}) \rightarrow CM^{i}\ and\ \\ &m(s^{i}_{k}, U^{i}_{k}) \rightarrow \texttt{SYNC\_REPLY}^{i}_{k}, k \neq 0 \}
  \end{split}
\end{equation*}

The previous proof from \texttt{SYNC\_CHANNEL} can be extended over multiple channels to expand the dependency set and prove that the \DMA{} protocol indeed satisfies the barrier condition for the \texttt{SYNC\_ONE} barrier as well.
Like the dependency set, the pending set of a \texttt{SYNC\_ONE} barrier is also a union of individual pending sets from each upstream actor:

\begin{equation*}
  \begin{split}
    \mathcal{P}_{B_{one}} = \bigcup_{i=1}^{N} \{ m(s^{i}_{j}, U^{i}_{j}) \mid &CM^{i} \rightarrow m(s^{i}_{0}, U^{i}_{0})\ and\ \\ &\texttt{SYNC\_REPLY}^{i}_{k} \rightarrow m(s^{i}_{k}, U^{i}_{k}), k \neq 0 \}
  \end{split}
\end{equation*}

\SP{} sent by the upstream actors $U^{i}_{0}$ consist of \texttt{DEPENDENCY\_PAYLOAD} message that instructs each downstream instance $D_{k}$ to block incoming messages at a particular Sequence ID on a channel.
Note that the Sequence ID mentioned in \texttt{DEPENDENCY\_PAYLOAD} only buffers messages on a particular channel. Messages from other upstream actors $U^{j}$ shall still be received and processed until the \SP{} from $U^{j}$ is received. Eventually, all upstream actors in $B_{one}$ shall send their $CM$, and the barrier will be complete.